\documentclass[lettersize,journal]{IEEEtran}
\usepackage[colorlinks=true, linkcolor=blue, citecolor=blue, urlcolor=magenta]{hyperref}
\usepackage{amsmath,amsfonts}
\usepackage{algorithm}
\usepackage{algpseudocode}
\usepackage{array}
\usepackage[caption=false,font=normalsize,labelfont=sf,textfont=sf]{subfig}
\usepackage{textcomp}
\usepackage{stfloats}
\usepackage{url}
\usepackage{graphicx}
\usepackage{verbatim}
\usepackage{epstopdf}
\usepackage{dsfont}
\usepackage{amssymb}
\usepackage{amsthm}
\newtheorem{theorem}{Theorem}  

\newtheorem{remark}{Remark}

\newtheorem{lemma}{Lemma}

\usepackage{booktabs} 
\usepackage{cite}
\usepackage{booktabs}
\usepackage{tikz}
\usetikzlibrary{arrows.meta}
\usepackage{fontawesome5}
\usetikzlibrary{arrows.meta,positioning,calc,fit,backgrounds,shapes.geometric,decorations.pathreplacing}
\usetikzlibrary{arrows.meta,decorations.pathreplacing}

\begin{document}

\title{MDP-based Energy-aware Task Scheduling\\ for Battery-less IoT}

\author{Shahab~Jahanbazi,~\IEEEmembership{Student Member,~IEEE},
        Mateen~Ashraf,
        Onel~L.~A.~L\'opez,~\IEEEmembership{Senior Member,~IEEE}
        
\thanks{This work is partially supported in Finland by the Research Council of Finland (Grants 362782 (ECO-LITE), and 369116 (6G Flagship)); and by the European Commission through the Horizon Europe/JU SNS project AMBIENT-6G (Grant 101192113).\\
The authors are with the Centre for Wireless Communications, University of Oulu, 90014 Oulu,
Finland (e-mail: \{shahab.jahanbazi; mateen.ashraf; onel.alcarazlopez\}@oulu.fi).}}


\maketitle

\begin{abstract}
Battery-less Internet of Things (IoT) devices rely on ambient energy harvesting and therefore require scheduling policies that jointly account for energy intermittency and hard timing constraints. This challenge is especially acute in periodic monitoring applications, where a sensing--computing--transmitting task chain must be completed within each reporting cycle. In this paper, we formulate this problem within a setting characterized by independently and identically distributed (i.i.d.) energy arrivals as a long-term average-reward Markov decision process (MDP) that explicitly captures capacitor-voltage evolution, task ordering, permissible start windows, and safe-execution requirements. We further propose rewards that promote reliable task completion while penalizing risky low-energy execution. We prove that the considered MDP is unichain and that the optimal stationary policy has a threshold structure, which leads to an optimal stationary threshold-based (OSTB) scheduler. To account for more realistic energy sources, we additionally study a correlated harvesting model based on a finite-state Markov process and show that the proposed framework can be applied to this richer setting under conservative sufficient conditions. Finally, numerical results show that OSTB outperforms representative baselines in terms of long-term full-chain completion rate, power failures, and latency, particularly when harvested energy is scarce.
\end{abstract}

\begin{IEEEkeywords}
Battery-less IoT devices, Markov decision process, task scheduling, energy harvesting.
\end{IEEEkeywords}

\section{Introduction}
\IEEEPARstart{B}{attery}-less Internet of Things (IoT) is a key enabler for sustainable large-scale sensing, since it can reduce maintenance costs, avoid battery replacement, and support long-lived operation~\cite{sustainable,onelnewpaper}. These devices harvest energy from ambient sources, making device operations inherently intermittent. Energy-aware protocols mitigate this behavior by allowing to dynamically adjust device schedules based on available energy \cite{delgado2022optimal,sabovic2020lora}. They are mainly meant for task scheduling \cite{moser2007realtime,sabovic2020lora,sabovic2022scheduler,rao2015optimal, 9174941,hao2019robust} and/or checkpointing~\cite{Hicks2017clank,8715130} processes.

Checkpointing involves periodically or adaptively saving system states to ensure forward progress, albeit at the cost of additional energy and time overheads~\cite{Hicks2017clank}. Meanwhile, task scheduling aims at efficiently allocating atomic tasks to computing resources over time. These approaches are often complementary, since task scheduling mitigates the risk of power failure, whereas checkpointing ensures that progress can be preserved and restored across power failure cycles~\cite{kang2024powerdepletion}. Note that in battery-less IoT devices that execute collections of atomic tasks, checkpointing is not often necessary and/or affordable. For such systems, where long-running programs are uncommon, and memory and data-processing capabilities are limited, task scheduling emerges as a more suitable and effective strategy. Importantly, task execution decisions should depend on the instantaneous energy state, as energy availability directly impacts system operation~\cite{moser2007realtime,sabovic2020lora,sabovic2022scheduler,delgado2022optimal,hao2019robust,rao2015optimal}.

Task scheduling becomes particularly important in periodic monitoring applications comprising a sequence of operations within each reporting cycle~\cite{periodic_sensing,sabovic2020lora}. For instance, the sequence may consist of sensing, local computing, and wireless transmission, which must be executed in order, and the sensing result remains useful only if the entire chain is completed within the intended cycle.
This makes timing constraints as important as energy constraints because excessive delay reduces data freshness and may violate application requirements. At the same time, premature task initiation can be equally harmful, as the device may experience a power failure before the ongoing task is completed. Therefore, an effective scheduler must reason jointly about \emph{when} a task is still admissible, \emph{whether} it can be completed safely, and \emph{how} the current decision affects future opportunities within the same cycle and across subsequent cycles.
The resulting scheduling problem is thus governed by a fundamental tradeoff. This is, waiting longer may allow more energy to be harvested and reduce the risk of power failure, but it also shrinks the remaining permissible execution window and may leave insufficient time for the downstream tasks. This tradeoff is challenging to tame, especially because the incoming energy is random and potentially temporally correlated. These observations motivate the need for a stochastic control framework that can explicitly account for energy uncertainty, task ordering, and hard timing constraints, while optimizing performance over a long operational horizon.

\subsection{Related Work}
Existing work on task scheduling for battery-less IoT devices can be broadly grouped into four categories: prediction-driven approaches, stochastic energy-aware formulations, intermittent real-time execution frameworks, and learning-based or application-specific control methods.

In the first category, scheduling decisions rely on known or estimated future energy availability~\cite{moser2007realtime,sabovic2020lora,sabovic2022scheduler,delgado2022optimal}. An early example is presented in~\cite{moser2007realtime}, where task execution is postponed to preserve stored energy. In the finite-power case, however, the latest safe starting time is computed using both the current energy level and the energy expected to be harvested before the deadline. More recent battery-less schedulers in~\cite{sabovic2020lora, sabovic2022scheduler, delgado2022optimal} account for task dependencies, device states, and capacitor dynamics, and achieve strong practical performance. However, they focus on finite-horizon or per-cycle formulations and require measured, profiled, or predicted energy traces.

The second category addresses \emph{stochastic energy uncertainty} more explicitly~\cite{hao2019robust,rao2015optimal,kang2024powerdepletion, biason2018decentralized}. The formulation in~\cite{hao2019robust} models uncertainty in energy harvesting (EH) for sensing and transmission services, while the Markov decision process (MDP) framework in~\cite{rao2015optimal} optimizes average utility over a long horizon under stochastic task and energy arrivals. More detailed stochastic energy-management formulations also appear in~\cite{kang2024powerdepletion, biason2018decentralized}. These works treat deadlines and priorities as soft utility terms and focus on energy-management abstractions.

The third category focuses on \emph{intermittent real-time execution}~\cite{karimi2021realtime,islam2020deadlineaware,karimi2022energyadaptive,karimi2026cartos}. The schedulability-oriented framework in~\cite{karimi2021realtime} studies periodic real-time tasks on intermittently powered battery-less devices. The authors in \cite{islam2020deadlineaware} jointly schedule computation and EH tasks in deadline-aware intermittent systems, while~\cite{karimi2022energyadaptive} further incorporates lightweight energy prediction to adapt sensing decisions according to future charging conditions. More recently, CARTOS \cite{karimi2026cartos} extends this direction to a charging-aware real-time operating system with support for task chains, mixed preemption, and just-in-time checkpointing. These studies are highly relevant from a systems perspective, but their emphasis is schedulability, freshness, or runtime support rather than long-term average optimal scheduling with an analytically characterized policy structure.

The fourth category includes \emph{learning-based or application-specific battery-less control}. Energy-aware tinyML task management in \cite{sabovic2023tinyml,jahanbazi2026multiexit} considers battery-less inference pipelines and application tasks. Reinforcement-learning-based systems such as ACES \cite{fraternali2020aces}, SmartON \cite{luo2021smarton}, and Ember \cite{fraternali2020ember} learn to adapt sensing or duty cycling under uncertain EH and event patterns. These methods are attractive when detailed models are unavailable, but they mainly target sensing quality, event capture, or application-level adaptation, and do not provide an analytically tractable scheduling policy for a periodic chain with hard execution windows.

\subsection{Contributions}
This paper studies energy-aware task scheduling for a battery-less IoT device executing a periodic task chain under strict deadline and energy constraints. Instead of focusing only on short-term feasibility~\cite{delgado2022optimal}, single-cycle optimization~\cite{sabovic2020lora,sabovic2022scheduler}, or simplified energy/timing models~\cite{kang2024powerdepletion,biason2018decentralized,rao2015optimal}, we formulate the scheduling problem as a long-term stochastic optimization problem using an MDP that captures capacitor-state evolution, execution feasibility, and within-cycle timing structure. The main contributions are summarized as follows.

\begin{enumerate}
\item \textbf{Hard-constrained MDP formulation for periodic battery-less task execution in i.i.d.-EH case.}
We formulate periodic task scheduling for a battery-less IoT device that repeatedly executes a fixed sensing--computing--transmitting chain, and cast the problem as a finite-state MDP under i.i.d.-EH case. Compared with prior energy-aware formulations in which deadlines or timing preferences are mainly captured through utility or penalty terms, e.g.,~\cite{rao2015optimal, kang2024powerdepletion, biason2018decentralized}, our formulation explicitly enforces hard permissible start windows, task precedence constraints, and safe-execution conditions induced by capacitor-voltage dynamics. Moreover, unlike methods that optimize decisions over a single cycle or a short prediction horizon, e.g.,~\cite{delgado2022optimal}, the proposed MDP incorporates capacitor voltage, intra-period timing, and task-chain progress into the system state, enabling long-term average optimization. 
\item \textbf{Reward design for conservative energy-safe policies.}
We design reward functions that couple scheduling decisions with the probability of safe task completion. Compared with learning-based energy-management approaches that adapt sensing or duty-cycling decisions according to energy availability, e.g.,~\cite{fraternali2020aces,fraternali2020ember}, and reward-design studies for EH sensor-node control, e.g.,~\cite{rioual2021design}, the proposed sigmoid-based reward explicitly promotes conservative scheduling in intermediate-energy states. Its tunable parameters allow the scheduler to reduce power failures while still preserving the long-term objective of maximizing completed task chains.
\item \textbf{Structural characterization of the optimal policy.}
We establish key analytical properties of the proposed hard-constrained MDP. Prior EH scheduling works have used MDP formulations for task execution and transmission control, e.g.,~\cite{rao2015optimal,biason2018decentralized}, while recent battery-less scheduling frameworks have focused on MILP-based optimization, dependency-aware scheduling, or real-time intermittent execution, e.g.,~\cite{delgado2022optimal,sabovic2022scheduler,karimi2021realtime}. In contrast, we characterize the structure of the optimal policy for the proposed periodic task-chain formulation. First, we show that the induced Markov chain (MC) is unichain under the stated conditions. Second, we prove that the optimal stationary policy has a threshold-based structure with respect to the capacitor voltage. This result reduces the policy search from a general state-action mapping to a set of voltage thresholds indexed by the within-period execution stage. Following these, we derive the optimal stationary threshold-based (OSTB) scheduler.
\item \textbf{Extension of OSTB to correlated EH.}
To address the practical limitation of the i.i.d. harvested-current assumption, we extend the formulation to a correlated EH model by augmenting the state with a finite-state EH mode governed by an MC. This extension preserves the scheduling logic of the original framework while allowing temporally correlated energy arrivals to be represented in a tractable way.
\item \textbf{Comprehensive evaluation against representative baselines.}
We evaluate the proposed scheduler under both i.i.d. and correlated EH models and compare it against representative baselines: an earliest-deadline-first-style (EDF) heuristic~\cite{Chetto2023EDF,Ghor2025DeadlineDriven} with an energy guard (EG) and an as-late-as-possible (ALAP) policy~\cite{moser2007realtime}. The comparison is carried out in terms of long-term task completion rate, execution latency, and number of power failures. This evaluation is intended to show not only that OSTB improves performance, but also \emph{why} it does so: by jointly adapting to the instantaneous energy state and the remaining time within the permissible execution window.
\end{enumerate}

The remainder of this paper is organized as follows. Section~\ref{sec:system_model} introduces the system model, Section~\ref{sec:formulation} formulates the problem, and Section~\ref{sec:Proposed_mdp} defines the MDP components. In Section~\ref{sec:optimal_policy}, we outline the objective function, discuss the foundational properties of the policy, analyze the characteristics of the optimal policy, and derive the resulting optimal scheduling strategy. Section~\ref{sec:correlated_case} extends the proposed framework to a correlated EH model that reflects real-world scenarios. Section~\ref{sec:baselines} presents two baselines.
Section~\ref{sec:simulation} presents simulation results, and Section~\ref{sec:conclusion} concludes the paper while including directions for future research.

\section{System Model}
\label{sec:system_model}

\begin{figure*}[!t]
\centering
\vspace{-1mm}

\resizebox{.96\textwidth}{!}{%
\begin{tikzpicture}[
    x=1cm,y=1cm,
    >=Latex,
    font=\scriptsize,
    panel/.style={draw, rounded corners=5pt, line width=.7pt},
    task/.style={draw, rounded corners=2pt, minimum height=.42cm, fill=gray!15, align=center},
    sleep/.style={draw, minimum height=.25cm, fill=gray!8},
    every node/.style={inner sep=1pt}
]
\draw[panel] (0,0) rectangle (15,2.65);
\draw[->, thick] (.45,1.18) -- (14.55,1.18);
\node[below] at (.45,1.05) {$0$};
\node[below] at (14.55,1.15) {$T_s$};
\node[above right] at (14.45,1.18) {time};
\foreach \x in {.45,1.10,...,14.15}
    \draw (\x,1.08)--(\x,1.30);
\node[anchor=east] at (14.25,2.42) {one cycle};
\draw[decorate, decoration={brace, amplitude=3pt}, thick]
    (.45,1.95)--(3.7,1.95);
\node[above=4pt] at (2.12,1.95)
    {$\mathcal{PTW}^{(s)}=\{0,\ldots,d_s\}$};
\draw[decorate, decoration={brace, amplitude=3pt}, thick]
    (3.1,1.5)--(10.2,1.5);
\node[above=4pt] at (6.90,1.5)
    {$\mathcal{PTW}^{(c)}=\{n_s,\ldots,M-n_c-n_t\}$};
\draw[decorate, decoration={brace, amplitude=3pt}, thick]
    (5.70,2)--(12.13,2);
\node[above=4pt] at (8.48,2)
    {$\mathcal{PTW}^{(t)}=\{n_s+n_c,\ldots,M-n_t\}$};
\draw[dashed] (1.10,.72)--(1.10,1.62);
\node[above] at (1.18,1.58) {$z_k^{(s)}\Delta t$};
\draw[dashed] (3.05,.72)--(3.05,1.92);
\draw[dashed] (3.70,.72)--(3.70,2.2);
\node[above] at (3.8,2.22) {$d_s\Delta t$};
\draw[dashed] (5.65,.72)--(5.65,2.22);
\node[task, minimum width=1.95cm] at (2.079,.78) {sensing};
\draw[<->] (1.10,.43)--(3.1,.43);
\node[below] at (2.18,.43) {$n_s\Delta t$};
\node[sleep, minimum width=1.25cm] at (3.7,.78) {};
\node[below=0pt] at (3.7,.65) {$q_k^{(s,c)}\Delta t$};
\node[task, minimum width=1.30cm] at (5,.78) {computing};
\draw[<->] (4.36,.43)--(5.66,.43);
\node[below] at (5,.43) {$n_c\Delta t$};
\node[sleep, minimum width=2.57cm] at (6.98,.78) {};
\node[below=0pt] at (7.15,.65) {$q_k^{(c,t)}\Delta t$};
\node[task, minimum width=3.20cm] at (9.9,.78) {transmitting};
\draw[<->] (8.3,.43)--(11.45,.43);
\node[below] at (10.08,.43) {$n_t\Delta t$};
\draw[<->] (12.78,1.4)--(13.45,1.4);
\node[above] at (13.1,1.50) {$\Delta t$};
\node[sleep, minimum width=2.58cm] at (12.8,.78) {};
\node[below] at (13.08,.60) {residual sleep until next cycle};
\end{tikzpicture}%
}

\vspace{-2mm}
\caption{One reporting period of duration $T_s$, during which the device executes a \emph{sensing} $\rightarrow$ \emph{computing} $\rightarrow$ \emph{transmitting} task chain subject to hard permissible start windows, precedence constraints, and the requirement that transmission be completed before the end of the cycle.}
\label{fig:system_model_tikz}
\end{figure*}

The device harvests ambient energy through an EH transducer, conditions it via a power management unit (PMU), and temporarily stores it in a capacitor that powers the sensing, computing, and transmission modules. The device operates in periodic reporting cycles of $T_s$ seconds in which each cycle consists of a task chain \emph{sensing} $\rightarrow$ \emph{computing} $\rightarrow$ \emph{transmitting}. Such a task chain is representative of modern low-power IoT operation, where sensed data may first undergo lightweight processing, feature extraction, compression, or event detection before being transmitted. More specifically, at the beginning of each cycle, the device tries to initiate a sensing task during which a target/environment parameter is measured. If the available energy is insufficient, sensing may be postponed until a predefined deadline~$d_s$. This bounded sensing-start window captures practical scheduling and wake-up delays while ensuring data freshness and timely operation \cite{KIM2025107542, age-based10.1145/3209582.3209602}.
Once sensing is completed, the device must first perform local processing on the sensed data and then report the result to a monitoring center. The resulting data must be fully transmitted within the ongoing cycle. Each cycle is further divided into~$M$ equal sub-intervals/time units of length $\Delta t = T_s/M$ (cf.~\ref{fig:system_model_tikz}). This subdivision enables a finer-grained characterization of the system's behavior and facilitates the evaluation of its performance within each cycle. Following this, the durations of the sensing, computing, and transmission operations are introduced as $n_s$, $n_c$, and $n_{t}$ units, respectively.

To characterize the execution process, we define the \emph{permissible time window} for each task $i \in \mathcal{I} \triangleq \{s,c,t\}$, denoted by $\mathcal{PTW}^{(i)}$. The indices $i=s$, $i=c$, and $i=t$ correspond to the active modes \texttt{sensing}, \texttt{computing}, and \texttt{transmitting}, respectively.
The set $\mathcal{PTW}^{(i)}$ specifies the time units at which the execution of task $i$ must start. 
As illustrated in Fig.~\ref{fig:system_model_tikz}, the permissible time windows are given by
\begin{align}
    \mathcal{PTW}^{(s)} &= \{0,\cdots,d_s\}, \nonumber\\
    \mathcal{PTW}^{(c)} &= \{n_s,\cdots,M-n_c-n_t\}, \nonumber\\
    \mathcal{PTW}^{(t)} &= \{n_s+n_c,\cdots,M-n_t\}.
\end{align}
Note that the computing task can only start after the sensing task has been executed and must still leave sufficient time for the subsequent transmission task to be completed within the ongoing main interval. 
Indeed, the earliest starting time of computing is $n_s$, since computing can begin only after the sensing task has occupied its required $n_s$ time units from the beginning of the interval.
Similarly, the transmission task can begin only after both sensing and computing have been completed, while remaining within the current main interval.
The device may also sleep with a one-sub-interval duration and fixed energy consumption during idle periods. This basic fine-grained low-power mode inclusion ensures that no potential operational scenario is overlooked. In particular, when the available energy in a sub-interval is inadequate to perform any task, modeling sleep allows the system to wait for the next opportunity to initiate execution.

We assume that the harvested current during the $n$-th sub-interval, denoted by $\mathbf{i}_h(n)$, is constant throughout the considered sub-interval and is assumed to be i.i.d. across the sub-intervals. Moreover, the connected load behaves as a constant resistance \cite{delgado2022optimal,sabovic2020lora,onelnewpaper} for each mode. Specifically, each mode is modeled by a load resistance $R_i = \frac{E}{I_i}$, where $E$ is the operating voltage and $I_i$ denotes the current drawn in mode $i \in \mathcal{I}\cup\{l\}$. The index $i=l$ denotes the \texttt{sleeping} mode.
During both sleep and task execution modes, the available energy in the capacitor undergoes variation due to discharging or charging. 
Assuming that the initial capacitor voltage at the start of the $n$-th sub-interval is $v_0$, and the device enters a certain mode $i$ for $n_{i}$ sequential sub-intervals, the resulting capacitor voltage after executing the ongoing task is given by
\begin{align}
\label{eq:finalvoltage}
 & \mathbf{v}_{i,n}(v_0,n_i)= e^{-\frac{n_{i}\Delta t}{R_{i} C}}\times \bigg( v_0+\nonumber\\
 & \quad \quad \quad  R_{i}(1-e^{-\frac{\Delta t}{R_{i} C}})\sum_{j=0}^{n_{i}-1}\mathbf{i}_h(n+j)e^{(j+1)\frac{\Delta t}{R_{i} C}} \bigg).
\end{align}
Note that for the special case of $n_{l} = 1$, this transforms into the well known
$e^{-\frac{\Delta t}{R_{l} C}}( v_0+R_{l}(1-e^{-\frac{\Delta t}{R_{l} C}})\mathbf{i}_h(n)e^{\frac{\Delta t}{R_{l} C}})$~\cite{delgado2022optimal}.

One key phenomenon that affects the decision-making process is the potential occurrence of a power failure. More specifically, the device cannot complete the execution of the ongoing task if the capacitor voltage drops below the turn-off threshold $V_\mathrm{out}$. Due to this, we introduce \textit{safe-execution probability} metric for the execution of task~$i$, defined as
\begin{equation}
\label{eq:vsafedef}
P_{\mathrm{safe}}^{(i)}(v_0) \triangleq \mathbb{P}(\mathbf{v}_{i,n}(v_0, n_i) \geq
V_{\mathrm{out}}), \quad \forall i\in\mathcal{I}.
\end{equation}
It is worth noting that, since the harvested current is assumed to be i.i.d., this parameter is independent of $n$, and therefore we drop this from the notation of $P_{\mathrm{safe}}^{(i)}(v_0)$. Moreover, $n_i$ is omitted, since the task identifier itself encodes the execution time, which is given as a prefix.
The device must execute task~$i$ within $\mathcal{PTW}^{(i)}$, during which the safe-execution probability is high. All in all, efficient task scheduling must comprehensively capture all the aforementioned restrictions and constraints.

\section{Problem Formulation}
\label{sec:formulation}
We focus on maximizing the long-term average number of successfully executed tasks. To formalize this, let $z_k^{(i)}$ and $v_{z^{(i)}_k}$ denote the starting time unit and its related voltage to execute task $i$ during the $k$-th main interval, respectively. Formally,
\begin{align}
   & z^{(s)}_k \in \mathcal{PTW}^{(s)},\nonumber\\
   & z^{(c)}_{k} = z^{(s)}_k+n_s+q_k^{(s,c)} \in \mathcal{PTW}^{(c)},\nonumber\\
   & z^{(t)}_{k} = z^{(c)}_k+n_c+q_k^{(c,t)} \in \mathcal{PTW}^{(t)},
    \label{eq:premi_z_deif}
\end{align}
where $q_k^{(s,c)}$ and $q_k^{(c,t)}$ denote the number of optional sleep sub-intervals inserted between two consecutive tasks. Their feasible ranges are
\begin{align}
q_k^{(s,c)} &\in \{0,\ldots,M-n_c-n_t-(z_k^{(s)}+n_s)\}\triangleq \mathcal{Q}^{(c)}_k, \nonumber\\
q_k^{(c,t)} &\in \{0,\ldots,M-n_t-(z_k^{(c)}+n_c)\}\triangleq \mathcal{Q}^{(t)}_k. \nonumber
\end{align}
Furthermore, to measure forward progress, let us define the task completion indicator as
\begin{equation}
    \xi_k^{(i)}=\mathds{1}\{\exists z^{(i)}_k : P_{\mathrm{safe}}^{(i)}(v_{z^{(i)}_k})\geq1-\epsilon\}, \forall i\in \mathcal{I},
    \label{eq:num_objec}
\end{equation}
where $\mathds{1}\{.\}$ denotes the indicator function and $\epsilon$ is the maximum tolerable risk threshold.

\begin{figure*}[t]
\begin{equation}
v_{k,0}=
\begin{cases}
\mathbf{v}_{l,0}(v_{k-1,0},M), & \xi_{k-1}^{(s)}=\xi_{k-1}^{(c)}=\xi_{k-1}^{(t)}=0, \\
\mathbf{v}_{l,z_{k-1}^{(s)}+n_s}(\mathbf{v}_{s,z_{k-1}^{(s)}}(v_{z_{k-1}^{(s)}},n_s),M-z_{k-1}^{(s)}-n_s), &  \xi_{k-1}^{(s)}=1, \xi_{k-1}^{(c)}=\xi_{k-1}^{(t)}=0,\\
\mathbf{v}_{l,z_{k-1}^{(c)}+n_c}(\mathbf{v}_{c,z_{k-1}^{(c)}}(v_{z_{k-1}^{(c)}},n_c),M-z_{k-1}^{(c)}-n_c), &  \xi_{k-1}^{(s)}=\xi_{k-1}^{(c)}=1, \xi_{k-1}^{(t)}=0,\\
\mathbf{v}_{l,z_{k-1}^{(t)}+n_{t}}(\mathbf{v}_{t,z_{k-1}^{(t)}}(v_{z_{k-1}^{(t)}},n_{t}),M-z_{k-1}^{(t)}-n_{t}), &  \xi_{k-1}^{(s)}=\xi_{k-1}^{(c)}=\xi_{k-1}^{(t)}=1.
\end{cases}
\label{eq:v_zero}
\end{equation}
\noindent\rule{\textwidth}{0.4pt}
\end{figure*}

Let $v_{k,0}$ denotes the capacitor voltage at the beginning of the $k$-th main interval, then using $\xi_{k-1}^{(i)}$ we can write $v_{k,0}$ in the recursive form shown in \eqref{eq:v_zero}. Note that once the final possible task has been executed, the device will remain in sleep mode for the remainder of the interval. 
The first case in \eqref{eq:v_zero} represents the scenario where no task is executed during the previous main interval, and the device remains in sleep mode throughout that interval. For the second case in \eqref{eq:v_zero}, if only sensing is executed, then the device enters sleep mode once the task is completed and stays in that mode until the end of the interval (i.e., from $z_{k-1}^{(s)}$ to the $M$-th time unit). Similarly, in the third case, the device enters sleep mode after completing the computing task. Likewise, if the transmitting task is executed, as in the fourth case in \eqref{eq:v_zero}, the device transitions to sleep mode immediately after completing the task and remains in that state for the remainder of the interval.

The voltage at time unit $z_k^{(s)}$ depends on $v_{k,0}$ as
\begin{equation}
    v_{z_{k}^{(s)}}=
    \begin{cases}
        v_{k,0}, & z_{k}^{(s)}=0, \\
        \mathbf{v}_{l,0}(v_{k,0},z_{k}^{(s)}), & z_{k}^{(s)}\in \mathcal{PTW}^{(s)}\setminus 0,
    \end{cases}
    \label{eq:vfin_def_sen}
\end{equation}
where the second case corresponds to the final voltage after the device has remained in sleep mode for $z_k^{(s)}$ units.
Furthermore, the capacitor voltage at time unit $z_k^{(j)}$ to start performing task $j\in \{c,t\}$ within the $k$-th main interval can be written as
\begin{align}
    v_{z_{k}^{(j)}}=
    \begin{cases}
        \mathbf{v}_{j,z_k^{(j)}}(v_{z_k^{(j)}},n_j), & q_k^{(a_j,j)}=0, \\
        \mathbf{v}_{l,z_k^{(j)}+n_j}(\mathbf{v}_{j,z_k^{(j)}}(v_{z_k^{(j)}},n_j),q_k), & q_k\in \mathcal{Q}^{(j)}_k\setminus 0,
    \end{cases}
    \label{eq:vfin_def_tx}
\end{align}
where $a_j=s$ for $j=c$ and $a_j=c$ for $j=t$. The first case represents the capacitor voltage immediately after completing the previous task, while the second case corresponds to the voltage after the device has remained in sleep mode for $q_k$ time units following the executed task.

The optimization problem is formulated as
\begin{subequations}\label{eq:optgeneral}
\begin{align}
    & \underset{\{v_{z^{(s)}_k},v_{z^{(c)}_k},v_{z^{(t)}_k}\}}{\text{maximize}} \quad  \lim_{K\to \infty} \frac{1}{K} \sum_{k=1}^{K} \bigg(w_s \xi_k^{(s)}+w_c\xi_k^{(s)}\xi_k^{(c)}\nonumber\\ 
    & \quad\quad\quad\quad\quad\quad\quad\quad\quad\quad\quad\quad\quad\quad\quad +w_t\xi_k^{(s)}\xi_k^{(c)}\xi_k^{(t)}\bigg) \label{eq:optob}   \\
    & \text{subject to}  \quad  v_{z_{k}^{(i)}} \in[V_{min},V_{max}], \quad \forall i\in\mathcal{I} , k\in \mathbb{N},   \\
    & \quad\quad\quad\quad\quad \eqref{eq:premi_z_deif}, \eqref{eq:num_objec}, \eqref{eq:v_zero}, \eqref{eq:vfin_def_sen}, \eqref{eq:vfin_def_tx},
\end{align}
\end{subequations}
\noindent
where $V_{min}$ and $V_{max}$ correspond to the minimum and maximum permissible operating voltages, respectively. Furthermore, $w_s,w_c,w_t\ge 0$ are weighting parameters. By choosing $(w_s,w_c,w_t)=(1,1,1)$, the objective rewards partial forward progress along the task chain, whereas $(w_s,w_c,w_t)=(0,0,1)$ yields a pure end-to-end success criterion. Therefore, \eqref{eq:optob} provides a flexible way to balance intermediate processing gains and final reporting success.
Since~\eqref{eq:optgeneral} involves stochastic constraints and its objective function is defined over a long-term time horizon, directly tackling this problem is not straightforward. To address this, we reformulate it in the next section as an MDP, which offers a structured framework for modeling long-term sequential decision-making under uncertainty. Indeed, we incorporate the aforementioned constraints into the MDP components' design and leverage the framework to facilitate the search for an effective solution and to derive the values of $\{v_{z^{(s)}_k},v_{z^{(c)}_k},v_{z^{(t)}_k}\}_{k\in \mathbb{N}}$ systematically.

\section{MDP Design}
\label{sec:Proposed_mdp}
In this section, we first define the key components of the MDP-based approach. We then introduce two transition matrices, which together allow for a more tractable analytical representation. Finally, we present two distinct reward functions that form the basis of the proposed objective function.

\subsection{State Space}
\label{section:state}
Since the execution time for sensing, computing, and transmitting tasks comprises several sub-intervals, we define an \textit{epoch} in the MDP framework as a single decision stage, corresponding to one transition in the state space (i.e., observation of the current state, action selection, and transition to the next state).
At the $n$-th epoch, the state of the system is captured by a three-element vector $s_n$ as
\begin{equation}
    s_n=(V_n,\tau_n,f_n),
\label{eq:definstate}
\end{equation}
where $V_n\in[V_{min},V_{max}]$ denotes the capacitor voltage, $\tau_n\in \{0,\cdots,M-1\}$, referred to as the \textit{local clock}, represents the index of the sub-interval within the main interval. Finally, the \textit{task flag} $f_n\in \{0,1,2,3\}$ indicates whether sensing, computing, or transmitting tasks have already been executed in the current main interval. Specifically, $f_n=0$ if the sensing task is executable, $f_n=1$ and $f_n=2$ indicate that the sensing and computing tasks, respectively, have already been completed within the current main interval, and $f_n=3$ denotes that the transmission task has already been performed within the ongoing interval.

As the voltage variable $V_n$ is continuous, the associated state space is infinite. To enable the application of classical finite-state MDP methods and to facilitate the evaluation of long-term performance criteria, we quantize $V_n$ onto $N_v$ levels, i.e., $V_n \in \mathcal{V}\triangleq \{V^{(1)},V^{(2)},\cdots,V^{(N_v)}\}$, where $V_{min}=V^{(1)}<V^{(2)}<\cdots<V^{(N_v)}=V_{max}$. At the beginning of each main interval, both the task flag and the local clock are reset to $\tau_n=f_n=0$. Following this, the set of all admissible system states forms the state space~$\mathcal{S}$, which contains a finite number of elements, denoted by $N_s$.

With the above state space design, the next state of the process depends solely on the current state and the action taken in that state, rather than the entire history of past states and actions. This property clearly indicates that the problem adheres to the non-memory Markov property, which is fundamental in MDPs.

\subsection{Action Space}
\label{sec:action}
The action space of the considered system is defined as
\begin{align}
    \mathcal{A} = \{\texttt{sleeping},\,& \texttt{sensing},\nonumber\\ 
     &\texttt{computing},\, \texttt{transmitting}\}.
\end{align}
At the beginning of each sub-interval, the device observes its current state $s = (V, \tau, f)$ where the variables $\tau$ and $f$ jointly determine whether a specific task remains feasible within the current interval.
To formalize this, we define the subsets of the state space as
\begin{align}
\mathcal{S}_1 &= \{(V,\tau,f) : \tau \in \mathcal{PTW}^{(s)},\ f = 0\},\nonumber\\
\mathcal{S}_2 &= \{(V,\tau,f) : \tau \in \mathcal{PTW}^{(c)},\ f = 1\},\nonumber\\
\mathcal{S}_3 &= \{(V,\tau,f) : \tau \in \mathcal{PTW}^{(t)},\ f = 2\}.
\end{align}
Consequently, sensing, computing, and transmission tasks can only be executed when the system state satisfies $s \in \mathcal{S}_1$, $s \in \mathcal{S}_2$, and $s \in \mathcal{S}_3$, respectively.
Given the current state $s$, the device selects an action $a$ from the set of admissible actions $\mathcal{A}_s$, defined as
\begin{equation}
    \mathcal{A}_{s} =
    \begin{cases}
        \{\texttt{sleeping},\ \texttt{sensing}\}, & s \in \mathcal{S}_1, \\
        \{\texttt{sleeping},\ \texttt{computing}\}, & s \in \mathcal{S}_2, \\
        \{\texttt{sleeping},\ \texttt{transmitting}\}, & s \in \mathcal{S}_3, \\
        \{\texttt{sleeping}\}, & \text{otherwise},
    \end{cases}
    \label{eq:actionset}
\end{equation}
which is state-dependent but time-invariant.

\subsection{State Transition Probability Matrix}
The state transition probability matrix captures the stochastic evolution of the system by defining the transition probabilities between states conditioned on the selected actions.
Note that the assumption of i.i.d. EH process across sub-intervals leads to a transition matrix that remains invariant with respect to the process's evolution.
For ease of readability, we initially consider the deterministic part of the state (i.e., $\tau$ and $f$) and obtain $N_v$ states for each possible pair, and denote it as superstate $(\tau, f)$.
More formally, the collection of states corresponding to the fixed parameters $\tau$ and $f$ is referred to as a \textit{superstate}~$(\tau,f)$, including $N_v$ states as $(V,\tau,f), \forall ~ V \in \mathcal{V}$.

Following the division of state into two parts, we define two levels of transition matrices to model the system dynamics: micro transition matrices, which capture transitions for individual actions, and a macro transition matrix, which integrates these micro transition matrices across the entire state space.

\subsubsection{Micro Transition Matrices}
The transitions between voltage levels, occurring as the system evolves from one superstate to another, are governed by a set of action-dependent transition matrices. Each of these matrices is of size $N_v \times N_v$ and encodes the probabilities of transitioning between discrete voltage levels under a specific action. 
As there are different possible actions, we define four micro transition matrices, denoted by $A_{l}$, $A_s$, $A_c$, and $A_{t}$, which correspond respectively to the \texttt{sleeping}, \texttt{sensing}, \texttt{computing}, and \texttt{transmitting} actions.

\subsubsection{Macro Transition Matrix}
The macro transition matrix is a matrix of size $N_s \times N_s$, constructed from the micro transition matrices $A_{l}$, $A_s$, $A_c$, $A_{t}$, and the zero matrix $\mathbf{0}_{N_v \times N_v}$. It encodes all transitions between all possible superstates in the system. Specifically, the matrix has the following properties:

\begin{itemize}
    \item \textbf{\texttt{Sensing} action:} For superstate $(\tau\in \mathcal{PTW}^{(s)},f=0)$ the next is $(\tau+n_s,f=1)$ which is governed by $A_{s}$.
    \item \textbf{\texttt{Computing} action:} For superstate $(\tau\in \mathcal{PTW}^{(c)},f=1)$ the next is $(\tau+n_c,f=2)$ which is governed by $A_{c}$.
    \item \textbf{\texttt{Transmitting} action:} For superstates $(\tau\in \mathcal{PTW}^{(t)}, f = 2)$
    \begin{itemize}
        \item if $\tau < M-n_{t}$, the next superstate is $(\tau + n_{t}, f = 3)$,
        \item if $\tau = M- n_{t}$, the next superstate is $(0, 0)$.
    \end{itemize}
    This transition is governed by $A_{t}$.
    \item \textbf{\texttt{Sleeping} action:} Governed by $A_{l}$, which advances the system by one sub-interval. For superstates $(\tau, f)$ where:
    \begin{itemize}
        \item $0 \leq \tau < M - 1$, $f = 0$,
        \item $n_s \leq \tau < M - 1$, $f = 1$,
        \item $n_s + n_{t} \leq \tau < M - 1$, $f = 2$,
    \end{itemize}
    the next superstate is $(\tau + 1, f)$.
    \item \textbf{Interval reset:} At the end of the main interval, the system resets the local clock and task flag. For superstates $(\tau=M-1,f\in \{0,1,2,3\})$, the next superstate is $(0, 0)$, with transition governed by $A_{l}$.
    \item \textbf{Invalid transitions:} For all other superstate pairs not covered above, the transition matrix is defined as $\mathbf{0}_{N_v \times N_v}$.
\end{itemize}

\begin{figure*}[!t]
    \centering
    {
        \includegraphics[width=0.5\textwidth]{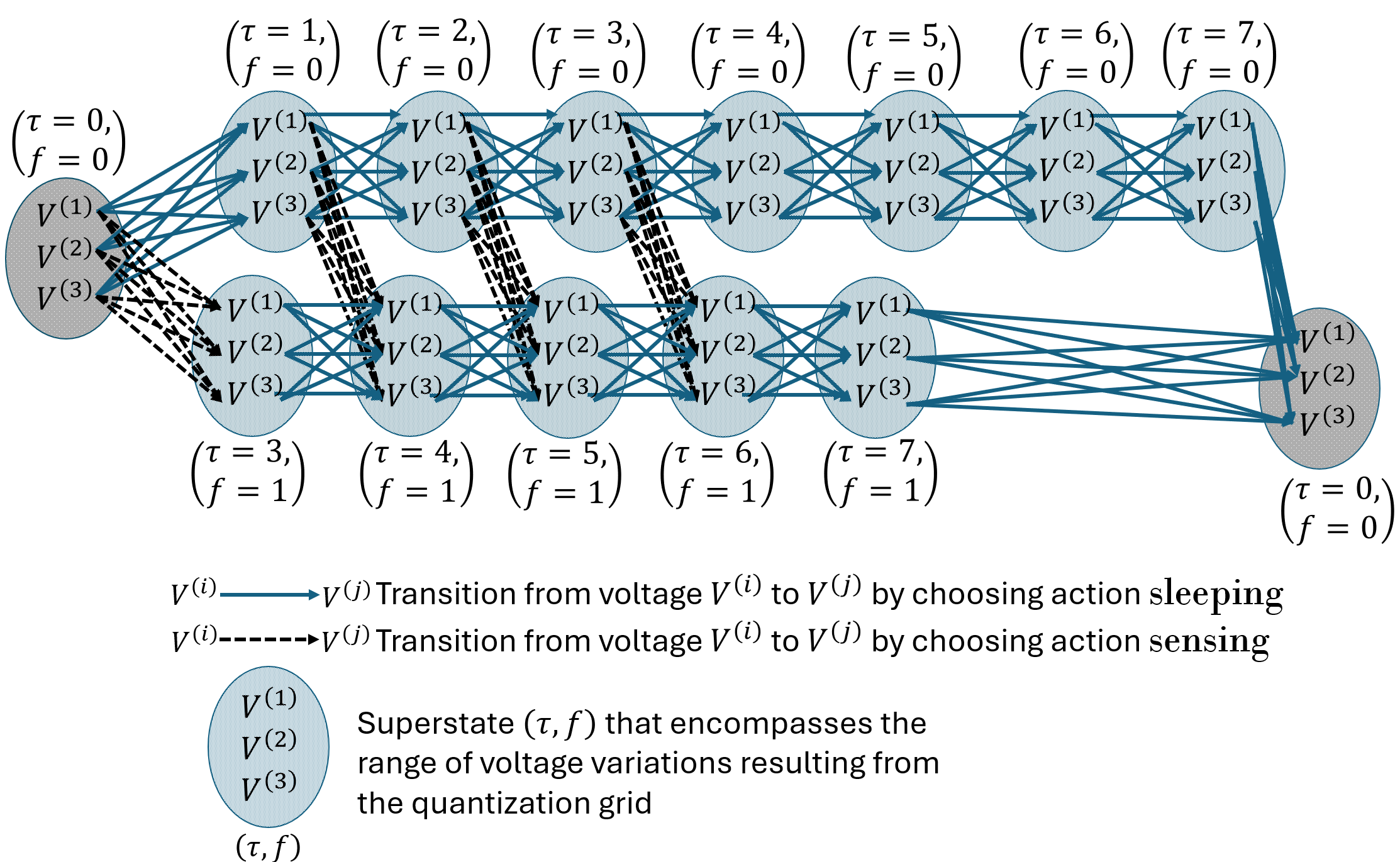}
        \label{fig:forwardprogress}
    }
   {
        \includegraphics[width=0.47\textwidth]{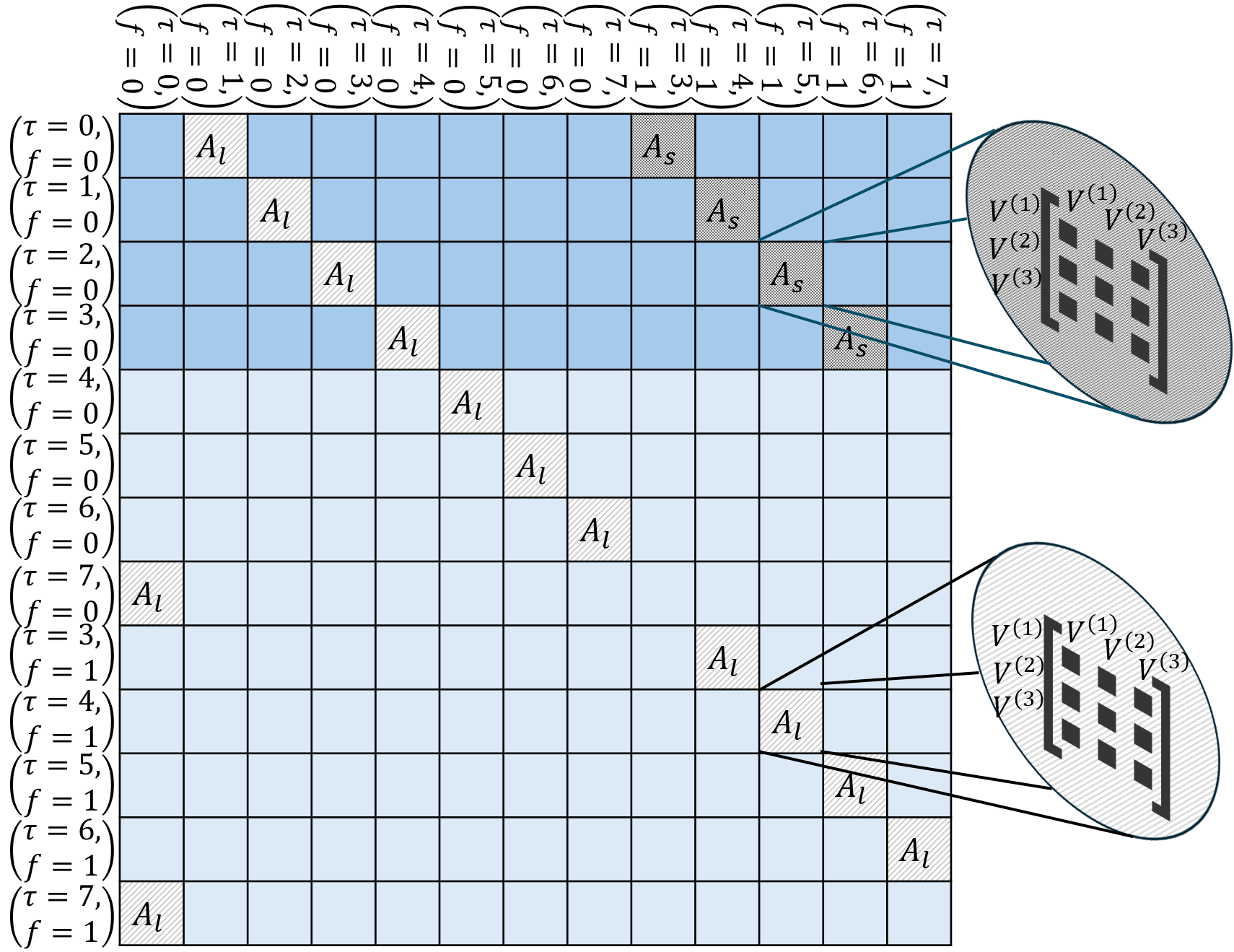}
        \label{fig:generalmatrix}
    }
    \caption{An illustrative representation of the scheme described in Section~\ref{ex:superstates}, assuming $M=8$, $N_v=3$, $d_s=3$, and $n_s=3$. Fig.~\ref{fig:superstate} (left) shows the forward progress of the process, including the defined superstates. Fig.~\ref{fig:superstate} (right) presents the general form of the transition matrix, where all possible action choices are considered. Each part in this matrix corresponds to a sub-matrix associated with a specific action that leads to a transition to another state in a different superstate. In particular, for rows corresponding to superstates with $(\tau\in\{0,1,2,3 \},f=1)$, the policy selects the appropriate rows from either $A_{l}$ or $A_s$, depending on the action taken. Light and dark blue boxes denote zero matrices.}
    \label{fig:superstate}
\end{figure*}

\subsubsection{Example}
\label{ex:superstates}
Here, we provide an example to illustrate the above concepts. Without loss of generality, we assume that the device can take only two actions: \texttt{sleeping} and \texttt{sensing}. However, the discussion in the example can be naturally extended to include more actions as well.

Consider a periodic approach in which the sensing task is executed once every $T_s=1$ second. Moreover, every one-second interval is divided into $M=8$ equal sub-intervals. The sensing task is characterized by two parameters: a deadline of $d_s=3$ sub-interval units, and an execution time of $n_s=3$ units. The voltage quantization grid is defined with $N_v=3$ discrete levels. As shown in Fig. \ref{fig:superstate} (left), to satisfy the deadline constraint, the agent for each state within the first four superstates $(\tau\in \{0,1,2,3\},f=0)$ can choose \texttt{sleeping} or \texttt{sensing}. These choices are closely tied to the selected policy, which is discussed in detail in Section~\ref{sec:optimal_policy}. More specifically, the system can transition from state $(V^{(i)},\tau,0)$ where $\tau\leq 3$ to state $(V^{(j)},\tau+n_s,1)$ with probability $A_{s}(i,j)$, provided the agent selects \texttt{sensing} action. Alternatively, if the agent chooses \texttt{sleeping} action in state $(V^{(i)},\tau,f)$, where $f\in\{0,1\}$, the system will move to state  $(V^{(j)},\tau+1,f)$ for $\tau<M-1$ and to state $(V^{(j)},0,0)$ for $\tau=M-1$ with probability $A_{l}(i,j)$.
    
Furthermore, in Fig.~\ref{fig:superstate} (right), the overall structure of the macro transition matrix is depicted, capturing all possible actions that may lead from one state to another. Note that the action selected by the agent, as determined by the policy, is reflected only in the first four (i.e., $d_s+1$) rows of the transition matrix, corresponding to the superstates $(\tau\in \{0,1,2,3\},f=0)$. For states belonging to the remaining superstates, the only permissible action is \texttt{sleeping}.

\subsection{Reward Function}
\label{sec:reward_defin2}
We design two reward functions tailored to reflect the system's operational efficiency, thereby ensuring that only safe actions contribute to the agent’s performance in each state. They capture the reward received by the agent upon taking action $a$ in state $s=(V,\tau,f)$.

\subsubsection{Basic Reward Function}
This reward function is primarily defined in terms of the safe-execution probability as
\begin{equation}
r(s,a)=
\begin{cases}
0, & a=\texttt{sleeping},\\[2mm]
w_i\,P_{\mathrm{safe}}^{(i)}(V), & a\in\mathcal{A}_s\setminus \texttt{sleeping},
\end{cases}
\label{eq:rewarddefinn2}
\end{equation}
where $i\in\mathcal{I}$ indexes the task associated with action $a$ and $(w_s,w_c,w_t)\in\mathbb R_+^3$ are the same weighting parameters used in~\eqref{eq:optgeneral}.
This function assigns higher rewards to states with higher voltage levels, thereby encouraging the agent to actively engage in task execution when sufficient energy is available. Moreover, it is a non-decreasing and monotonic function with respect to the initial voltage (i.e., $V$), which can reduce the complexity of the MDP formulation \cite{puterman1994} as further discussed in Section~\ref{sec:thresholdpolicy}.

\subsubsection{Sigmoid-based Reward Function}
\begin{figure}
    \centering
    \begin{minipage}{0.22\textwidth}
        \centering
        \includegraphics[width=\linewidth]{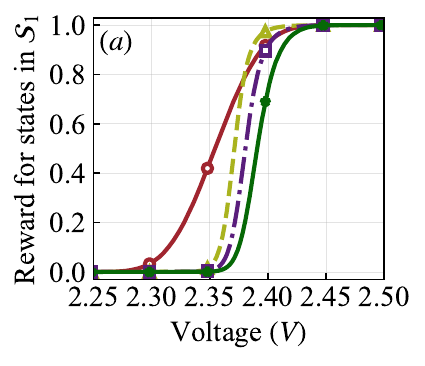}
    \end{minipage}
    \begin{minipage}{0.253\textwidth}
        \centering
        \includegraphics[width=\linewidth]{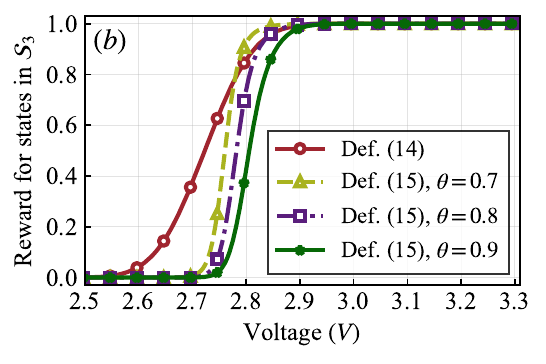}
    \end{minipage} 
    \caption{Reward values from \eqref{eq:rewarddefinn2} and \eqref{eq:rewarddefin2} with $\beta=15$ and $\theta \in \{0.7,0.8,0.9\}$ as functions of voltage for states in $\mathcal{S}_1$ with action $\texttt{sensing}$ (a) and states in $\mathcal{S}_3$ with action $\texttt{transmitting}$ (b). We set $C=1.7\text{mF}$, $V_{out}=2.4\text{V}$, and $\mathbf{i}_h$ as uniform i.i.d. between $0$ and $4$ mA, i.e., $\mathcal{U}[0,4]\text{mA}$.}
\label{fig:rewards_tasks}
\end{figure}

This reward function includes additional parameters to provide greater control over power failure. Specifically, we use a normalized sigmoid function, characterized by a steepness parameter $\beta$ and threshold value $\theta$ as
\begin{equation}
r(s,a)=
\begin{cases}
0, & a=\texttt{sleeping},\\[2mm]
w_i\,\frac{\sigma^{(i)}(V)}{\sigma^{(i)}(V_{max})}, & a\in\mathcal{A}_s \setminus \texttt{sleeping},
\end{cases}
\label{eq:rewarddefin2}
\end{equation}
where $\sigma^{(i)}(V)\triangleq1/(1+e^{-\beta(P_{\mathrm{safe}}^{(i)}(V)-\theta)})$ and $i\in\mathcal{I}$ denotes the corresponding task.
It should be noted that the parameter $\theta$ serves a restrictive role by effectively enforcing the scheduling of task $i$ whenever the value of $P_{\mathrm{safe}}^{(i)}(V)$ exceeds $\theta$. This mechanism effectively reduces the occurrence of power failures. More specifically, as illustrated in Fig. \ref{fig:rewards_tasks}, the reward values defined in \eqref{eq:rewarddefinn2} corresponding to states with moderate voltage levels are disregarded according to \eqref{eq:rewarddefin2}. Consequently, execution initiation occurs at higher voltage levels in the scheduling process, ultimately leading to a reduction in power failures. This benefit is further examined and discussed in detail in Section~\ref{subsec:OSTB}.

 \section{Optimal policy}
\label{sec:optimal_policy}
Herein, we first introduce the considered objective function, followed by the policy types in our MDP. In addition, we establish the unichainity property of the problem under consideration, and then show that the optimal policy has a threshold-based structure. Finally, we determine the optimal policy that maximizes the long-term average reward.

\subsection{Objective Function}
The objective is to develop a scheduling policy to maximize the task completion rate over a long-term horizon while minimizing the risk of power failure. To formally capture this objective, we adopt the long-term average reward criterion, a standard metric in MDP-based formulations. In this framework, the policy $\pi: \mathcal{S}\rightarrow \mathcal{A}_s$ is defined as a mapping from each state to its allowable action set.
Let $s_0 = s$ indicate the initial system state, then the average expected reward of a policy $\pi$, denoted by $g^\pi(s)$, is defined as
\begin{equation}
g^\pi(s)=\lim_{N\rightarrow \infty} \frac{1}{N} \mathbb{E} \left[ \sum_{n=0}^{N-1} r^\pi(s_n,a_n)\vert s_0=s \right],
\label{eq:objectfun} 
\end{equation}
where $r^\pi(s_n, a_n)$ denotes the reward received at the $n$-th epoch upon taking action $a_n$ in state $s_n$ under policy~$\pi$. Then, the aim is to find an optimal policy $\pi^*$ that \textit{maximizes the average expected reward} across all admissible policies as
\begin{equation}
\pi^* = \arg\max_{\pi} \, g^\pi(s),  \forall s \in \mathcal{S}.
\end{equation}
Since the average expected reward function is defined as a limit and depends on the initial state, determining the optimal policy requires a deeper analysis of its structural properties, which is presented in the following subsection.

\subsection{Policy Types and Unichainity Property}
MDP policies can be described along two dimensions: deterministic versus stochastic, and stationary versus non-stationary.
Among these, deterministic and stationary policies are of particular importance, as they often serve as natural candidates for both theoretical analysis and practical implementation~\cite{puterman1994}. Here, we examine these two properties in detail within the proposed MDP framework, highlighting their role in shaping decision-making across states and over time.

\textbf{Deterministic Policy:} A policy is said to be deterministic if, for each state $s\in\mathcal{S}$, it selects a single action with certainty \cite{puterman1994}. One common deterministic policy is threshold-based decision-making. For example, in the proposed MDP's states where sensing is permitted (i.e., $s\in\mathcal{S}_1$), the policy may choose the \texttt{sensing} action if the safe-execution probability satisfies $P_{\mathrm{safe}}(V)\geq \gamma$, and choose \texttt{sleeping} otherwise. Here, $\gamma$ is a fixed threshold parameter that may depend on the second component of the state (e.g., $\tau$). For instance, in Section \ref{ex:superstates}, a possible deterministic policy may be defined as follows. In the superstate $(\tau=2,f=0)$, the policy selects the action \texttt{sleeping} for all states of the form $(V,2,0)$, where $V\in\{V^{(1)},V^{(2)}\}$, and selects the action \texttt{sensing} for the state $(V^{(3)},2,0)$. In contrast, within the superstate $(\tau=3,f=0)$, the policy assigns the action \texttt{sleeping} to state where $V^{(1)}$, and the action \texttt{sensing} to states with $V\in \{V^{(2)},V^{(3)} \}$.

\textbf{Stationary Policy:} A policy is stationary if it remains unchanged over time and depends only on the current state, not explicitly on the time index or decision step \cite{8107579, puterman1994}. For example, in Section \ref{ex:superstates}, whenever the process occupies a state of the form $s=(V,\tau,f)$ with $\tau\in \{0,1,2,3\}$ and $f=0$, the agent applies the same transition rule to determine the subsequent state.

Certain structural properties of the MCs induced by policies can be exploited for analyzing MDPs with long-term average reward criteria to find the optimal policy. Specifically, for any deterministic stationary policy $\pi$, the induced MC partitions the state space into one or more irreducible recurrent classes, along with a possible set of transient states \cite{ross}. Moreover, according to \cite[Proposition 8.2.1]{puterman1994}, any two states within the same closed, irreducible recurrent class share the same average expected reward $g^{\pi}$. Consequently, if the MC induced by every deterministic stationary policy contains exactly one recurrent class (possibly with transient states), known as the unichain property, then $g^{\pi}(s)$ is the same for all states $s$. This ensures the optimality criterion is independent of the initial state and is robust across the entire state space.

\begin{theorem}
Let $\pi$ be an arbitrary stationary deterministic policy, and let the transition matrices $A_{l}$, $A_s$, $A_c$, $A_t$ be time-invariant and have strictly positive elements. 
Then, the MC induced by $\pi$ contains exactly one recurrent class (possibly along with a set of transient states), and thus the corresponding MC is unichain.
\label{th:unichain}
\end{theorem}

\renewcommand{\qedsymbol}{}
\begin{proof}
The proof is provided in Appendix \ref{app:th1}. 
\end{proof}

\subsection{Threshold-based property of the optimal policy}
\label{sec:thresholdpolicy}
The complexity of finding the optimal policy in an MDP stems from the size of the state and action spaces, as the search for the optimal policy must be conducted over their joint space.
Even without reducing the number of state-action pairs, the complexity of solving an MDP can be mitigated by designing a well-structured reward function that guides the policy towards simpler but more efficient forms, such as threshold-based decisions. This property enables efficient computation of the optimal policy by limiting the policy search to those with a threshold-based structure\cite{puterman1994}.

A threshold-based policy within our MDP refers to a decision rule in which, at each sub-interval within the permissible time window, the device chooses to execute the corresponding task only if the capacitor voltage exceeds a specified threshold. 

\begin{theorem}
\label{th:threshold}
Under the definition of MDP and both reward functions provided in Section \ref{sec:Proposed_mdp}, there exists an optimal deterministic stationary policy $\pi^*$ that is threshold-based in $V$ within each superstate. That is, there exist thresholds $\Gamma^{(\gamma(f))}_{m}$ such that, for all
$f\in\{0,1,2\}$ and $m\in\mathcal{PTW}^{(\gamma(f))}$,
\[
\pi^\star((V,m,f))=
\begin{cases}
\texttt{sleeping}, & V<\Gamma^{(\gamma(f))}_{m},\\
a(f), & V\ge \Gamma^{(\gamma(f))}_{m},
\end{cases}
\]
where $\gamma(0)=s$, $\gamma(1)=c$, $\gamma(2)=t$, and
$a(0)=\texttt{sensing}$, $a(1)=\texttt{computing}$, $a(2)=\texttt{transmitting}$.

\end{theorem}

\renewcommand{\qedsymbol}{}
\begin{proof}
The proof is provided in Appendix \ref{app:th2}. 
\end{proof}
\renewcommand{\qedsymbol}{$\square$} 

For unichain MDPs, the maximum long-term average reward and the corresponding optimal stationary policy can be computed using well-established methods, such as linear programs (LPs)~\cite{puterman1994,8107579}. Herein, formulate it as
\begin{align} 
& \underset{x(s,a)}{\text{maximize}}  \quad \sum_{s\in\mathcal{S}}\sum_{a\in\mathcal{A}_s} x(s,a)r(s,a) \nonumber\\
& \text{s.t.}  \quad  x(s,a)\geq 0 , \quad \forall s\in\mathcal{S} , a\in\mathcal{A}_s,   \nonumber\\
& \quad\quad \sum_{a\in\mathcal{A}_{s'}} x(s',a)=\sum_{s\in\mathcal{S}}\sum_{a\in\mathcal{A}_s} x(s,a)P(s'\vert s,a), \forall s'\in\mathcal{S},\nonumber\\
& \quad\quad \sum_{s\in\mathcal{S}}\sum_{a\in\mathcal{A}_s} x(s,a)=1.
\label{eq:optgeneral1}
\end{align}
As this is an LP, one can readily leverage convex optimization tools, like CVX, to compute the optimal stationary distribution $x^*(s,a)$ associated with the given MDP. According to \cite[Corollary 8.8.3]{puterman1994}, in any state $s\in \{\mathcal{S}_1,\mathcal{S}_2,\mathcal{S}_3\}$ where different actions are admissible, the resulting optimal policy prescribes selecting the action $a\in\mathcal{A}_s$ for which $x^*(s,a)>0$. This procedure directly determines the threshold values associated with the optimal scheduling strategy. The resulting policy is herein referred to as the OSTB scheduler.

\begin{remark}
   In the policy $\pi^*$, the thresholds $\Gamma_m^{(i)}$, for $i \in \mathcal{I}$, depend on $m$, representing the state component $\tau_n$. As long as the policy depends exclusively on the current state $s_n=(V_n,\tau_n,f_n)$ and not explicitly on the time index $n$, it is stationary. The apparent time dependence introduced through $\tau_n$ is incorporated into the state representation, and thus the policy remains a fixed mapping from states to actions and unchanged over time.
\end{remark}

\subsection{Computational Overhead and Scalability}
A practical advantage of the proposed OSTB policy is that its computational burden is largely \emph{offline}. In particular, LP in~\eqref{eq:optgeneral1} is solved offline to obtain the optimal stationary policy, whereas the device only executes the resulting threshold rule online. It is therefore important to distinguish between \emph{offline synthesis complexity} and \emph{online deployment complexity}.

The quantized i.i.d.-EH formulation has $N_s = O(MN_v)$ states, with the coarse upper bound $N_s < 4MN_v$.
The LP in~\eqref{eq:optgeneral1} uses the occupation measures $x(s,a)$ as decision variables. Since every state admits the \texttt{sleeping} action and only states in $\mathcal{S}_1$, $\mathcal{S}_2$, and $\mathcal{S}_3$ admit one additional task-execution action, the total number of LP variables is
\begin{equation}
N_x = \sum_{s\in\mathcal S}|\mathcal{A}_s| = N_s+N_vN_{\mathrm{th}},
\end{equation}
where $N_{\mathrm{th}}=|\mathcal{S}_1|+|\mathcal{S}_2|+|\mathcal{S}_3|$ is the number of thresholds.
The LP further contains $N_s$ flow-conservation equalities and one normalization equality, in addition to the nonnegativity constraints. Consequently, the offline optimization size is linear in the discretized state dimension.

Solving a densely constrained LP through interior-point method results in $O\!\left((MN_v)^3\right)$
up to logarithmic factors and solver-dependent constants. This bound is conservative, however, because the LP matrix induced by our MDP is highly sparse. In particular, each state-action pair transitions only to one successor superstate through one of the $N_v\times N_v$ micro transition matrices $A_l$, $A_s$, $A_c$, or $A_t$. As a result, each column of the constraint matrix contains only $O(N_v)$ nonzero entries, which makes sparse LP solvers substantially more efficient in practice than the dense worst-case bound suggests.

More importantly, the online implementation is much simpler than solving the MDP on-device. By Theorem~\ref{th:threshold}, the optimal policy is threshold-based in the capacitor voltage within each decision superstate. Therefore, at runtime, the device only needs to: i) identify the current superstate through $(\tau,f)$, ii) retrieve the corresponding threshold, and iii) compare the measured capacitor voltage against that threshold. Hence, the online decision complexity is $O(1)$ per sub-interval.
The memory footprint is also small because the deployed policy needs to store only $N_{\mathrm{th}}$ thresholds rather than the full state-action map.
Overall, the proposed framework is scalable for the compact periodic task-chain setting studied here, because the heavy computation is performed offline and grows polynomially with the discretized state dimension.

\section{Correlated EH Extension}
\label{sec:correlated_case}
EH from ambient sources often exhibits temporal dependence. Specifically, under slowly varying light or thermal conditions, the harvested current in one sub-interval is statistically informative about that in subsequent sub-intervals, so the i.i.d. assumption does not hold.
To incorporate such temporal dependence while preserving a tractable Markovian control formulation, we model the incoming harvested current through a finite-state Markov process. Specifically, we augment the original state $s_n=(V_n,\tau_n,f_n)$ by an EH mode variable $H_n$, and define the new state as
\begin{equation}
\hat{s}_n=(V_n,\tau_n,f_n,H_n),
\end{equation}
where $H_n\in \mathcal{H}\triangleq\{1,\dots,L\}$ represents the current harvesting mode with $L$ possible values. The permissible action sets remain unchanged. Thus, the correlated-EH extension preserves the scheduling logic of the original formulation.

\begin{lemma}
The augmented process $\{\hat{s}_n\}_{n\ge 0}$ is Markov.
\end{lemma}

\begin{proof}
Let $\mathbf P_H=[p_{hh'}]_{h,h'\in\mathcal H}$ denote the transition matrix of the EH-mode MC, where
\begin{equation}
p_{hh'}=\Pr(H_{n+1}=h' \mid H_n=h).
\end{equation}
Moreover, let mode $h$ correspond to a harvested current level~$I_h$. Then, under action $a\in\{l,s,c,t\}$ and a one-slot duration $\Delta t$, the capacitor voltage obeys the exact one-step RC update $V_{n+1}=e^{-\frac{\Delta t}{R_a C}} V_n + R_a(1- e^{-\frac{\Delta t}{R_a C}})I_{H_n}$, where $R_a=\frac{E}{I_a}$ is the load resistance associated with action~$a$. Hence, conditional on $(V_n,\tau_n,f_n,H_n)$ and the selected action, the pair $(V_{n+1},H_{n+1})$ is independent of the past. Therefore, the proof is completed.
\end{proof}

For a multi-slot task $i\in\mathcal{I}$ with duration $n_i$, the corresponding end-of-task voltage distribution is obtained by repeated composition of the one-step kernel together with the EH-mode transitions. Accordingly, the safe-execution probability corresponding to task $i\in\mathcal{I}$ becomes mode-dependent and is written as
\begin{equation}
P_{\mathrm{safe}}^{(i)}(v,h)
\triangleq\Pr\!\left(V_{n+n_i}\ge V_{\mathrm{out}}\mid V_n=v, H_n=h, a_n=i\right).
\end{equation}
Here, the dependence on the EH mode $h$ is explicit, in contrast to the i.i.d. model. Following this, the basic reward dedicated to state $\hat{s}=(V,\tau,f,H)$ by taking action $a$, can be written as
\begin{equation}
r(\hat{s},a)=
\begin{cases}
0, & a=\texttt{sleeping},\\[2mm]
w_i\,P_{\mathrm{safe}}^{(i)}(V,H), & a=\mathcal{A}_{\hat{s}}\setminus \texttt{sleeping},
\end{cases}
\end{equation}
where $i\in\mathcal{I}$ denotes the corresponding task.

Because the correlated setting is richer than the i.i.d. one, we take a conservative approach. Rather than asserting a fully general structural theorem for correlated EH dynamics, we state the following theorem, which provides sufficient conditions for threshold optimality in the augmented model.

\begin{theorem}
\label{theorem:non_iid}
Fix an EH mode $h\in\mathcal H$ and consider one of the three augmented superstate families
\[\hat {\mathcal S}^{(s)}_{m,h}=\{(V,m,0,h):V\in\mathcal V\},\qquad m\in \mathcal{PTW}^{(s)},\]
\[\hat {\mathcal S}^{(c)}_{m,h}=\{(V,m,1,h):V\in\mathcal V\}, \qquad m\in \mathcal{PTW}^{(c)},\]
\[\hat {\mathcal S}^{(t)}_{m,h}=\{(V,m,2,h):V\in\mathcal V\},\qquad m\in \mathcal{PTW}^{(t)}.\]
Suppose that for each fixed $(m,h)$, it holds that:

\emph{(i)} For every bounded increasing test function $\varphi$ on the next-state space, the difference
\begin{align}
V \mapsto {} & \mathbb{E}[\varphi(\hat{s}_{n+1}) \mid \hat{s}_n=(V,\tau,f,h), a\in\mathcal{I}] \nonumber\\
& - \mathbb{E}[\varphi(\hat{s}_{n+1}) \mid \hat{s}_n=(V,\tau,f,h), a=l]
\end{align}
is non-decreasing in $V$.

\emph{(ii)} The augmented MDP is unichain.

\noindent
Then there exists an optimal stationary policy that is threshold-based in the voltage variable within each augmented superstate. More precisely, define
\[(\alpha(f),a(f))=
\begin{cases}
(s,\texttt{sensing}), & f=0,\\
(c,\texttt{computing}), & f=1,\\
(t,\texttt{transmitting}), & f=2.
\end{cases}\]
Then, for each $h\in\mathcal H$, there exist mode-dependent thresholds
\[\bigl\{\Gamma^{(\alpha(f))}_{m,h}\bigr\}_{m\in\mathcal{PTW}^{(\alpha(f))}}, \qquad f\in\{0,1,2\},\]
such that, for all $f\in\{0,1,2\}$ and
$m\in\mathcal{PTW}^{(\alpha(f))}$,
\[\pi^\star((V,m,f,h))=
\begin{cases} \texttt{sleeping}, & V<\Gamma^{(\alpha(f))}_{m,h},\\
a(f), & V\ge \Gamma^{(\alpha(f))}_{m,h}.
\end{cases}\]

\end{theorem}

\begin{proof}
The proof is provided in Appendix \ref{app:non_iid}. 
\end{proof}

The above theorem provides sufficient, but not necessary, requirements for threshold-based policy to be optimal in correlated EH scenarios. In the correlated setting, verifying these conditions analytically for arbitrary $\mathbf P_H$ may be cumbersome. For this reason, we complement the theorem with a direct numerical verification tailored to the chosen correlated EH model. After solving the occupation-measure LP, we compute the gain-bias pair $(g,b)$ and evaluate the action-advantage functions defined in the proof of Theorem~\ref{theorem:non_iid},
\[
\Delta_s(V,m,h),\qquad
\Delta_c(V,m,h),\qquad
\Delta_t(V,m,h),
\]
for all states. If all these quantities are non-decreasing in $V$, then the corresponding policy is threshold-based in the discretized model. This numerical test is especially informative because it does not merely inspect the extracted policy itself, but verifies the stronger property that the underlying action preference changes monotonically with the capacitor voltage.

Regarding the computational overhead of this extension, the augmented state $\hat{s}_n$ increases the offline complexity linearly with the number of EH modes. Specifically, the number of states and actions for this case are given as
\begin{equation}
N_s^{\mathrm{corr}} = |\mathcal H|\,N_s,\qquad
N_x^{\mathrm{corr}} = |\mathcal H|\,N_x.
\end{equation}
On the other hand, the online complexity remains unchanged. The policy is still applied as a simple mode-dependent threshold lookup based on $(\tau,f,H)$. As a result, the runtime complexity stays constant at $O(1)$, while the memory scales as $O(N_{\mathrm{th}}|\mathcal{H}|)$.

\section{Baselines}
\label{sec:baselines}
\begin{figure*}[t]
\centering
\begin{tikzpicture}[
    scale=1,
    transform shape,
    x=1cm,y=1cm,
    >=Latex,
    font=\scriptsize,
    every node/.style={inner sep=1pt}
]
\draw[->, thick] (0,0) -- (12.4,0) node[right] {Time};
\node at (6.1,3.2) {\textbf{One main interval}};
\fill[gray!15] (0.5,2.9) rectangle (4,-0.4);
\fill[gray!8]  (4.25,2.9) rectangle (7.0,-0.4);
\fill[gray!25] (7.25,2.9) rectangle (11.5,-0.4);
\node at (2.0,-0.65) {$\mathcal{PTW}^{(s)}$};
\node at (5.5,-0.65) {$\mathcal{PTW}^{(c)}$};
\node at (9.5,-0.65) {$\mathcal{PTW}^{(t)}$};
\draw[dashed, gray] (0.5,-0.4) -- (0.5,3.0);
\draw[dashed, gray] (4,-0.4) -- (4,3.0);
\draw[dashed, gray] (4.25,-0.4) -- (4.25,3.0);
\draw[dashed, gray] (7.0,-0.4) -- (7.0,3.0);
\draw[dashed, gray] (7.25,-0.4) -- (7.25,3.0);
\draw[dashed, gray] (11.5,-0.4) -- (11.5,3.0);
\draw[dashed, gray] (1.6,-0.1) -- (1.6,0.8);
\draw[dashed, gray] (2.4,-0.1) -- (2.4,0.8);
\filldraw[black] (2.4,0) node[scale=1.2] {$\times$};
\filldraw[black] (1.6,0) node[scale=1.2] {$\times$};
\node[below] at (2.5,-0.1) {$t_0$};
\node[below] at (1.7,-0.1) {$t_0-\Delta t$};

\node[left] at (0,2.4) {{\Large\faClock}\,\textbf{ALAP}};
\node[left] at (0,1.5) {{\Large\faTachometer*}\,\textbf{EDF-EG}};
\node[left] at (0,0.6) {{\Large\faClock}\,\Large\faTachometer*\,\textbf{OSTB}};
\draw[thick] (0.2,2.4) -- (12,2.4);
\draw[thick] (0.2,1.5) -- (12,1.5);
\draw[thick] (0.2,0.6) -- (12,0.6);
\filldraw[black] (4,2.4) circle (2pt);
\filldraw[black] (7,2.4) circle (2pt);
\filldraw[black] (11.5,2.4) circle (2pt);
\node[above] at (3.9,2.4) {$s$};
\node[above] at (6.9,2.4) {$c$};
\node[above] at (11.4,2.4) {$t$};
\filldraw[black] (1.2,1.5) circle (2pt);
\filldraw[black] (4.7,1.5) circle (2pt);
\filldraw[black] (8.5,1.5) circle (2pt);
\node[above] at (1.1,1.5) {$s$};
\node[above] at (4.6,1.5) {$c$};
\node[above] at (8.4,1.5) {$t$};
\filldraw[black] (2.4,0.6) circle (2pt);
\filldraw[black] (5.4,0.6) circle (2pt);
\filldraw[black] (8,0.6) circle (2pt);
\node[above] at (2.3,0.6) {$s$};
\node[above] at (5.3,0.6) {$c$};
\node[above] at (7.9,0.6) {$t$};
\draw[->, thin] (3.3,3.2) -- (3.99,2.8);
\node[above] at (3.3,3.1) {$d_s\Delta t$};
\draw[->, thin] (1.7,1.85) -- (1.2,1.55);
\node[above] at (2.3,1.7) {$V\geq V_{\mathrm{th}}^{(s)}$};
\draw[->, thin] (5.2,1.85) -- (4.7,1.55);
\node[above] at (5.8,1.7) {$V\geq V_{\mathrm{th}}^{(c)}$};
\draw[->, thin] (9,1.85) -- (8.5,1.55);
\node[above] at (9.6,1.7) {$V\geq V_{\mathrm{th}}^{(t)}$};
\draw[->, thin] (2.6,1) -- (2.4,0.7);
\node[above] at (3.2,0.8) {$V\geq V_{\mathrm{th,t_0}}^{(s)}$};
\draw[->, thin] (0.5,1) -- (1.6,0.6);
\node[above] at (1.1,0.8) {$V< V_{\mathrm{th,t_0-\Delta t}}^{(s)}$};
\node[above] at (12.8,2) {$s$};
\node[align=left] at (13.8,2) {$\bullet$ \, Start of sensing};
\node[above] at (12.8,1.6) {$c$};
\node[align=left] at (13.95,1.6) {$\bullet$  \, Start of computing};
\node[above] at (12.8,1.2) {$t$};
\node[align=left] at (14,1.2) {$\bullet$  \, Start of transmitting};
\draw[thin] (12.6,2.3) -- (15.4,2.3);
\draw[thin] (12.6,2.3) -- (12.6,1);
\draw[thin] (12.6,1) -- (15.4,1);
\draw[thin] (15.4,2.3) -- (15.4,1);
\end{tikzpicture}
\caption{Illustrative comparison of task execution decisions under ALAP, EDF-EG, and OSTB over one main interval. ALAP schedules each task near the end of its permissible time window, EDF-EG executes at the earliest instant satisfying the voltage threshold condition, whereas OSTB adaptively balances safe execution and deadline satisfaction.}
\label{fig:policy_comparison}
\end{figure*}
Herein, we introduce two representative baseline policies to evaluate the proposed OSTB scheduling strategy. Both baselines operate under the same task ordering, allowable action sets, and execution constraints imposed by the system model. Fig.~\ref{fig:policy_comparison} provides an illustrative comparison of the task execution behavior under OSTB and the two baseline strategies.

\subsection{EDF-style with Energy Guard (EDF-EG)}
\label{subsubsec:EDF-EG}
We consider an EDF-EG heuristic~\cite{Chetto2023EDF,Ghor2025DeadlineDriven}. Under this policy, once a task becomes admissible, it is executed as early as possible to reduce the risk of deadline violation, provided that the available energy is sufficient to complete the task safely.
More precisely, for a given risk tolerance \(\epsilon \in (0,1)\), the scheduler first computes a voltage threshold \(V_{\text{th}}^{(i)}\) for each task according to
\begin{equation}
V_{\text{th}}^{(i)} \triangleq \min\left\{V \in \mathcal{V} : P_{\mathrm{safe}}^{(i)}(V) \geq 1 - \epsilon \right\}, 
\qquad i \in \mathcal{I}.
\label{eq:EDF-EG_threshold}
\end{equation}
Then, when the current state is $s=(V,\tau,f) \in \mathcal{S}_i$, the device executes task $i$ if and only if \(V \geq V_{\text{th}}^{(i)}\). In this sense, EDF-EG can be viewed as a policy primarily driven by the instantaneous energy state, in which the execution decision is determined solely by whether the capacitor voltage exceeds the task-specific threshold, without explicitly adapting to the remaining time within the permissible window.

\paragraph*{Choice of \(\epsilon\) and computation of \(P_{\mathrm{safe}}^{(i)}(V)\)}
The parameter \(\epsilon\) determines the tradeoff between reliability and aggressiveness. Smaller values of \(\epsilon\) yield a more conservative policy by requiring a higher probability of safe execution, whereas larger values allow more aggressive execution at the cost of an increased risk of power failure. For consistency, the same value of \(\epsilon\) is used here as in the proposed formulation. The quantity \(P_{\mathrm{safe}}^{(i)}(V)\) is estimated according to its definition in the system model via Monte Carlo sampling over EH realizations by checking whether the capacitor voltage remains above \(V_{\mathrm{out}}\) throughout the execution of task \(i\), given an initial voltage \(V\).

\subsection{As Late As Possible (ALAP)}
ALAP scheduling defers each task toward the end of its allowable time window. In contrast to EDF-EG, which primarily emphasizes energy sufficiency, ALAP is mainly driven by timing considerations. Its rationale is to leave as much time as possible for EH before task execution, thereby potentially improving the likelihood of safe completion.
We choose ALAP as a baseline for two main reasons. First, the scheduling problem addressed in this work involves only three tasks executed sequentially within \(T_s\)-length time slots, with task execution durations that are relatively short compared to \(T_s\). Under this structure, many recent scheduling heuristics developed for heterogeneous, parallel, or multiprocessor settings cannot be transferred directly without substantial modification. By contrast, ALAP can be implemented naturally within the present framework and therefore provides a simple and meaningful reference point. Second, because ALAP postpones execution until near the deadline, it reflects an intuitive reliability-oriented strategy in EH systems. In other words, harvesting for a longer period may reduce the risk of power failure by increasing the stored energy prior to execution.

At the same time, ALAP does not explicitly account for the risk that excessive postponement may leave insufficient time to complete subsequent tasks within the same main interval. This is particularly important in the present setting, where tasks must be executed in sequence and the EH is stochastic. Consequently, although ALAP may improve energy availability before executing a given task, it can also increase the likelihood of deadline violations for later tasks. This makes ALAP a relevant and informative benchmark: it captures the benefit of delayed execution for energy accumulation, while also highlighting the limitations of a policy that does not jointly balance energy safety and timing urgency.

\section{Simulation Results}
\label{sec:simulation}
In this section, we provide and analyze simulation results for both i.i.d. and correlated EH scenarios. For the former, we adopt a modeling approach consistent with prior studies on EH systems (e.g., \cite{9474499, 5934952mobile1}), where the harvested current is modeled as a uniform i.i.d. random process with uniform distribution.
For the latter, we consider a finite-state model characterized by a transition probability matrix that governs the evolution of the harvesting process across states.
Unless otherwise specified, all parameters are set according to Table~\ref{table:parameter}.

\begin{table}[t]
  \caption{Simulation parameters}
  \begin{tabular}{lll}
    \toprule
    \textbf{Description} & \textbf{Symbol} & \textbf{Value}\\
    \midrule  
    Turn-off threshold                   &  $V_\mathrm{out}$     & $1.8$ V  \\
    Capacitance                          &  $C$     & $\{0.7, 1.7, 2.7\}$ mF  \\
    Minimum Operating Voltage            &  $V_\mathrm{min}$     & $1.8$ V \\
    Maximum Operating Voltage            &  $V_\mathrm{max}$     & $3.3$ V \\
    Duration of Each Sub-interval        &  $\Delta t$    & $0.02$ s \\
    Deadline for executing sensing  &  $d_s\Delta t$ & $0.3$ s \\
    Number of sub-intervals              &  $M$           & $50$ \\
    Current Consumption of Sleep         &  $I_{l}$       & $0.1$ mA \\
    Current Consumption of Sensing  &  $I_{s}$       & $1.7$ mA \\
    Current Consumption of Computing      &  $I_{c}$  & $1$ mA \\
    Current Consumption of Transmitting  &  $I_{t}$  & $4.36$ mA \\
    Number of Levels for Descritization  & $N_v$          & $30$ \\
    Period Duration                      & $T_s$     & $1$ s \\
    Execution Time of Sensing       & $n_s\Delta t$  & $0.1$ s \\
    Execution Time of Computing     & $n_c\Delta t$  & $0.06$ s \\
    Execution Time of Transmitting   & $n_{t}\Delta t$& $0.4$ s \\
  \bottomrule
\end{tabular}
\label{table:parameter}
\end{table}

\subsection{Performance of the OSTB scheduling approach}
\label{subsec:OSTB}
\begin{figure}[!t]
    \centering
    \includegraphics[width=0.24\textwidth]{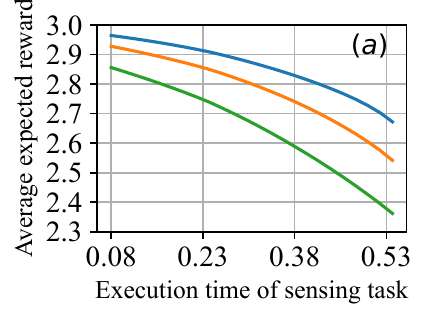}
    \includegraphics[width=0.24\textwidth]{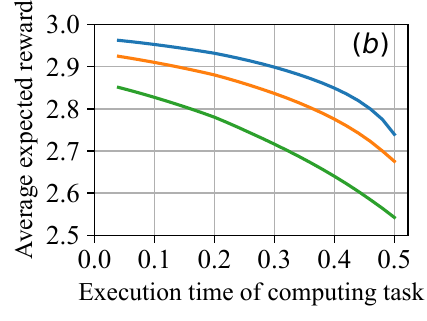}
    \includegraphics[width=0.36\textwidth]{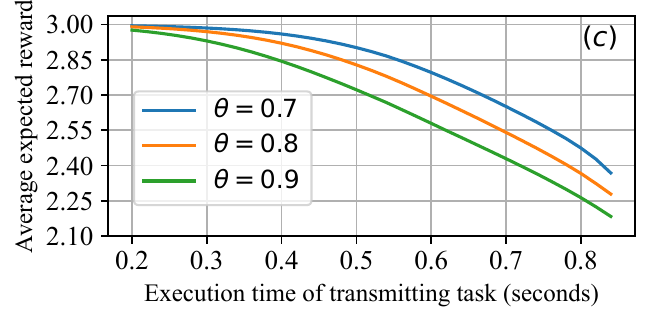}
    \caption{Average expected reward according to~\eqref{eq:objectfun} versus execution times of (a) sensing, (b) computing, and (c) transmitting. Results are shown for $\mathbf{i}_h \sim \mathcal{U}[0,4]$mA, $C=1.7$mF, $\beta = 25$ and $\theta \in \{0.7, 0.8, 0.9\}$.}
    \label{fig:combined}
\end{figure}

Fig.~\ref{fig:combined} shows the long-term average expected reward as a function of the execution time of each task, with the execution times of the other two tasks held constant, for three different values of $\theta$. Notably, the objective function values lie between $2$ and $3$, indicating that, on average, between two and three tasks are executed successfully per main interval.
It is shown that the objective function is decreasing and concave-like with respect to the execution time duration of tasks. This behavior arises from the reduced likelihood of completing subsequent tasks within their respective permissible time windows. For instance, in Fig.~\ref{fig:combined}~(c), as the transmitting execution time increases, the reduction in the average expected reward becomes more significant. This implies that the impact of increasing execution time is not linear: small increments in $n_t\Delta t$ at lower values (e.g., from $0.2$ to $0.3$ seconds) result in relatively minor changes in the objective, whereas the same increment at higher values (e.g., from $0.7$ to $0.8$ seconds) leads to a more substantial drop.

\begin{figure*}
  \centering
  \includegraphics[width=0.82\textwidth]{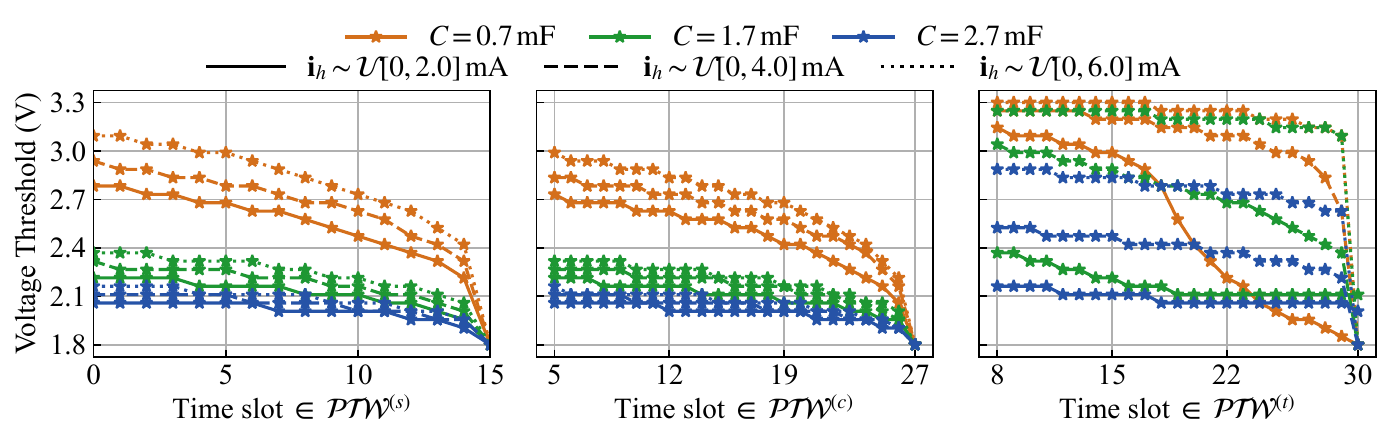}
  \caption{Threshold voltage versus the permissible time window to execute sensing (left), computing (middle), and transmitting (right), for $d_s=15$, $n_s=5$, $n_{c}=3$, $n_{t}=20$, and three distinct capacitor and EH statistics.}
  \label{fig:threshold_voltag}
\end{figure*}

By solving the LP in~\eqref{eq:optgeneral1}, we obtain the optimal stationary policy, which is fully characterized by a set of voltage thresholds over the permissible time windows of the tasks.
Fig.~\ref{fig:threshold_voltag} depicts these optimal threshold voltages for all task stages under different capacitance values and harvested-current distributions. The observed structure is fully consistent with Theorem~\ref{th:threshold}, which establishes that the optimal policy is threshold-based and therefore can be represented in a compact and interpretable form.

A first notable trend in Fig.~\ref{fig:threshold_voltag} is the temporal evolution of the thresholds within each permissible time window. At the beginning of a main interval, the policy is more conservative and typically requires a higher capacitor voltage before initiating execution. This behavior of deferring task execution in a low-energy state allows the device to harvest additional energy, thereby improving the probability of safe task completion. As the deadline approaches, however, the thresholds decrease progressively. In this regime, further postponement becomes less attractive because it increases the risk of missing the task deadline. Hence, near the end of the permissible window, the policy chooses to execute even at lower energy levels in order to preserve task feasibility. This temporal trend highlights the fundamental tradeoff captured by OSTB between execution safety and deadline compliance.

Fig.~\ref{fig:threshold_voltag} also reveals how the thresholds adapt to the energy environment and to the task characteristics. When the harvested current is more favorable, the optimal policy adopts higher thresholds, that is, it becomes more selective before starting a task. The reason is that under better EH conditions, postponement is less risky and can be exploited to accumulate a larger energy reserve, which in turn increases the probability of successful completion. In contrast, under poorer EH conditions, the policy lowers the thresholds to avoid excessive waiting that could eventually cause deadline violations.

Finally, the thresholds are not identical across the three task stages. The values in $\mathcal{PTW}^{(t)}$ are consistently higher than those in $\mathcal{PTW}^{(s)}$ and $\mathcal{PTW}^{(c)}$, which is explained by the larger execution time and higher current consumption of the transmission task. Because transmission is the most energy-demanding stage, safe execution requires a larger energy reserve, and the policy therefore acts more conservatively in $\mathcal{PTW}^{(t)}$. In addition, a smaller capacitance leads to larger threshold values. Although a smaller capacitor charges faster, it also discharges more rapidly and provides a smaller energy buffer, which increases the risk of interruption during task execution. The net effect is that the optimal policy requires a higher starting voltage in order to maintain safe operation. Overall, these results show that OSTB adapts its thresholds in a physically meaningful way to the remaining time, the task demand, and the energy-storage capability of the device.

\begin{figure}
  \centering
  \includegraphics[width=0.46\textwidth]{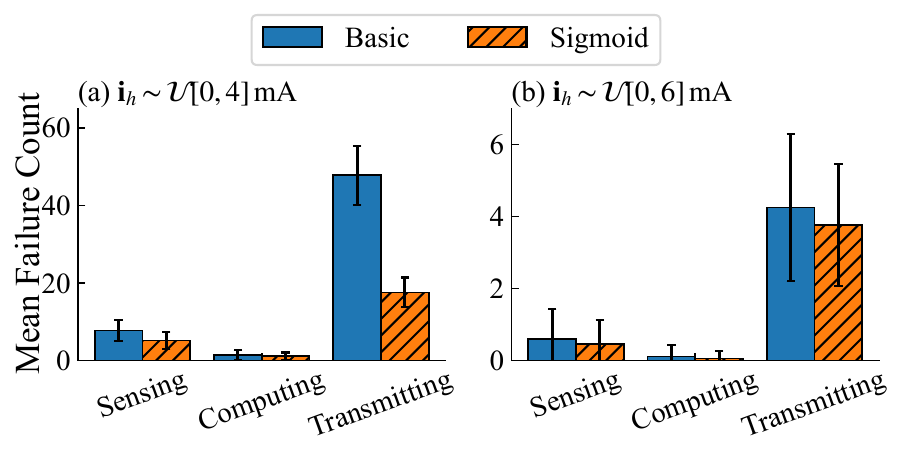}
  \caption{Mean and standard deviation of the number of failures following the execution of different tasks over a $1000$-s horizon, based on $100$ Monte Carlo simulations, for $C = 1.7$ mF.}
  \label{fig:comparing_rewards}
\end{figure}
To assess the impact of the reward choice, we solve~\eqref{eq:optgeneral1} under both reward definitions and evaluate the resulting policies over a $1000$-s horizon using $100$ Monte Carlo runs. Fig.~\ref{fig:comparing_rewards} reports the mean and std of the number of power failures observed after initiating the execution of the different tasks. As expected, power failures are more frequent when the EH is drawn from distributions with smaller support, since the scheduler operates more often in energy-limited conditions. More importantly, the figure clearly demonstrates the benefit of the sigmoid-based reward function. More specifically, it consistently yields fewer power failures than the basic reward across the considered operating regimes. This improvement stems from the fact that the sigmoid-based design penalizes low-energy execution states more sharply and rewards execution more selectively when the available energy is sufficiently high. As a result, the induced policy becomes more cautious in marginal energy conditions, avoiding risky task initiations that are likely to end in interruption, while still exploiting favorable energy states to maintain long-term task progress.

\subsection{OSTB vs Baselines}
\label{subsec:baselines}

\begin{figure}[!t]
    \centering
    \includegraphics[width=0.3\textwidth]{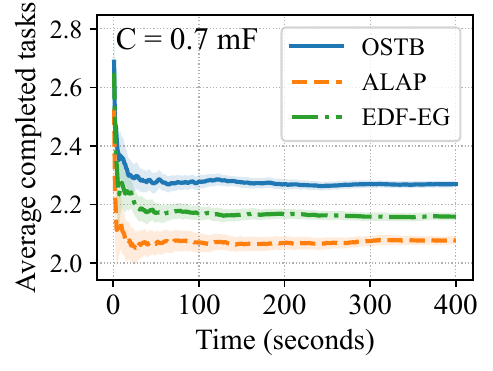}
    \includegraphics[width=0.3\textwidth]{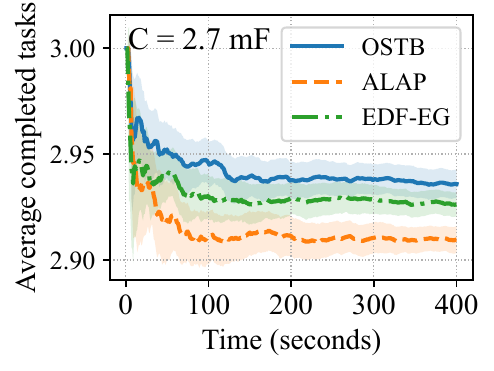}
    \caption{Average number of completed tasks per time unit. The harvested current is assumed to be $\mathbf{i}_{h}\sim \mathcal{U}[0,4]$ mA. The shaded regions represent the variability across $100$ Monte Carlo simulation runs.}
    \label{fig:Completed_task_conv}
\end{figure}
Fig.~\ref{fig:policy_comparison} depicts the working principle of EDF-EG, ALAP, and OSTP. Herein, we provide a comparison of the task execution performance of the proposed OSTB scheme and the baselines.

Fig.~\ref{fig:Completed_task_conv} presents the task completion rate per main interval over a $400$-s horizon, obtained from $100$ Monte Carlo simulations. It can be seen that OSTB consistently provides a higher task completion rate over the entire horizon. Within the considered system model, delaying task execution until the latest allowable instant may reduce the probability of completing all tasks within the interval. Owing to the stochastic EH nature, postponing the ``sensing'' task until close to its deadline may leave insufficient time or energy for the successful execution of the remaining tasks. This ultimately degrades the overall task completion rate.
In contrast, the EDF-EG policy does not explicitly account for the deadline constraint. Overall, ALAP and EDF-EG each neglect one of the key scheduling requirements, while OSTB simultaneously incorporates both deadline awareness and safe-execution considerations. This joint constraint consideration in OSTB results in its superior performance. Finally, the gap in task completion rate among ALAP, EDF-EG, and OSTB becomes smaller when the capacitance increases. With a larger capacitor, the voltage evolves more smoothly and becomes less sensitive to short-term EH fluctuations. Because the capacitor voltage directly drives scheduling decisions, the resulting performance becomes more stable and is slightly the same across the three policies. These results illustrate that OSTB is better suited for IoT devices with limited energy storage capacity.

\begin{figure}
  \centering
  \includegraphics[width=0.44\textwidth]{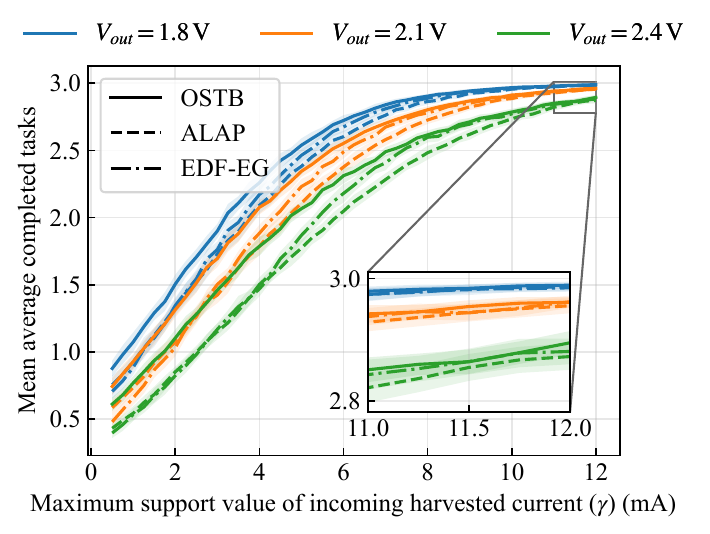}
  \caption{Average number of completed tasks per time unit versus different values of the support set of the harvested current over $1000$-second duration. The capacitance is assumed to be $C=0.7$mF. The shaded regions represent the variability across $100$ Monte Carlo simulation runs.}
  \label{fig:different_incoming_current_for__three_Vout}
\end{figure}
Fig.~\ref{fig:different_incoming_current_for__three_Vout} shows the task completion rate as a function of~$\gamma$, where the harvested current is modeled as $\mathbf{i}_h \sim \mathcal{U}[0,\gamma]$ over a $1000$-s interval, for different values of $V_{\mathrm{out}}$. The figure highlights the relative performance of the proposed OSTB and the baseline policies under both low and high harvested-current regimes. In particular, when the harvested current is limited (i.e., $\gamma < 7.5$ mA), OSTB exhibits a performance advantage. This indicates that the performance gains achieved by OSTB are more pronounced when the support range of the harvested current is narrow, thereby making OSTB especially suitable for operation under tightly energy-constrained conditions. By contrast, in the higher-current regime (i.e., $\gamma \geq 7.5$ mA), the completion rates of OSTB and EDF-EG converge to the same values. This behavior can be explained by the mean harvested current and the capacitor size. Since the capacitance is relatively small ($C = 0.7$ mF), the capacitor charges and discharges rapidly. Moreover, as shown in Fig.~\ref{fig:threshold_voltag}, the resulting threshold voltage remains relatively high and nearly constant over the permissible time window for these $\gamma$ and capacitor values, which becomes close to the threshold produced by EDF-EG in \eqref{eq:EDF-EG_threshold}. Consequently, in this region, OSTB effectively reduces to the EDF-EG policy.

\begin{figure*}[!t]
    \centering
    \includegraphics[width=0.32\textwidth]{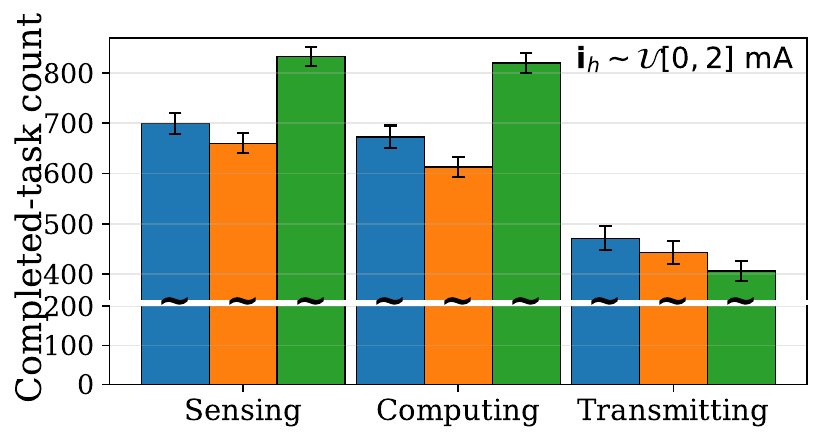}
    \includegraphics[width=0.32\textwidth]{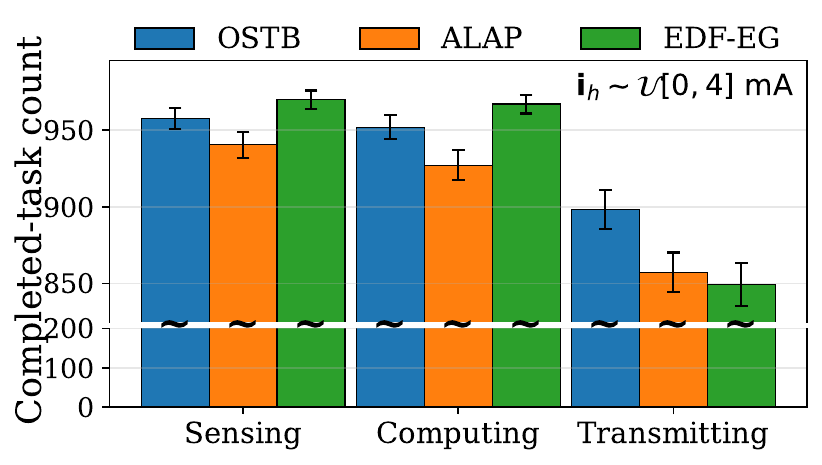}
    \includegraphics[width=0.32\textwidth]{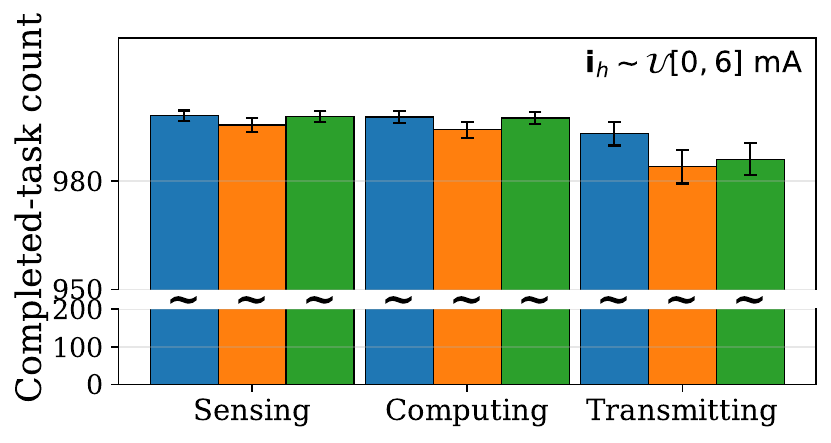}
    \caption{Average number of completed tasks per task for basic reward function definition. The capacitor is assumed to be $C=1.7$mF. The completed-task count is expressed as the mean and std derived from $100$ Monte Carlo simulations, each with a length of $1000$.}
    \label{fig:completed_tasks}
\end{figure*}
Fig.~\ref{fig:completed_tasks} highlights the long-term performance advantage of the proposed OSTB policy over the considered baselines. Over the $1000$-s evaluation horizon, OSTB yields a larger number of fully completed task cycles, as reflected by the higher number of completed ``transmitting'' tasks across the three incoming-current distributions. Recall that a transmission is counted as completed only when its two prerequisite tasks have also been successfully executed, and therefore, it provides the most meaningful indicator of end-to-end cycle completion. Interestingly, although OSTB may complete fewer individual ``sensing'' and ``computing'' tasks than EDF-EG, it allocates execution opportunities more effectively across the task chain, thereby converting a larger fraction of initiated cycles into fully successful ones. This gain is particularly pronounced under low incoming-current conditions, where energy scarcity makes myopic decisions more costly. The result confirms that OSTB better exploits the long-term tradeoff between immediate execution and future feasibility by jointly accounting for deadline constraints and energy availability.

\begin{table*}[t]
    \centering
    \caption{Latency performance comparison for the three-task sequential execution model.}
    \label{tab:delay_for_execution_time}
    \begin{tabular}{c ccc ccc ccc}
        \toprule
        & \multicolumn{3}{c}{$\mathbf{i}_h\sim \mathcal{U}[0,2]$ mA}
        & \multicolumn{3}{c}{$\mathbf{i}_h\sim \mathcal{U}[0,4]$ mA}
        & \multicolumn{3}{c}{$\mathbf{i}_h\sim \mathcal{U}[0,6]$ mA} \\
        \cmidrule(lr){2-4} \cmidrule(lr){5-7} \cmidrule(lr){8-10}
        & OSTB & ALAP & EDF-EG
        & OSTB & ALAP & EDF-EG
        & OSTB & ALAP & EDF-EG \\
        \midrule
        Mean latency (s)
        & 0.6149 & 1.0000 & 0.6333
        & 0.5966 & 1.0000 & 0.5747
        & 0.5748 & 1.0000 & 0.5604 \\
        Std of latency (s)
        & 0.0047 & 0 & 0.0062
        & 0.0050 & 0 & 0.0026
        & 0.0028 & 0 & 0.0003 \\
        \bottomrule
    \end{tabular}
\end{table*}

Latency is another key metric in sequential task scheduling. While it may appear desirable to execute sensing immediately at the beginning of each $T_s$-second interval and then start the remaining tasks as soon as possible, such behavior is not always energy-efficient in a battery-less system and may increase the risk of power failure. To quantify this effect, Table~\ref{tab:delay_for_execution_time} reports the average number of successful full-chain epochs together with the mean and standard deviation of execution latency for OSTB, ALAP, and EDF-EG over a $1000$-s horizon, assuming $C=2.7$ mF. Here, the latency is defined as the transmission completion time within a $1$-second epoch, and epochs affected by full-chain failure are excluded from the latency statistics. In the more energy-constrained regime, OSTB also yields a smaller mean latency than the baselines. 
In particular, for $\mathbf{i}_h\sim\mathcal{U}[0,2]$ mA, OSTB reduces the latency relative to both EDF-EG and ALAP, while maintaining the highest successful full-chain count (cf. Fig.~\ref{fig:completed_tasks}). The ALAP policy unsurprisingly yields a latency of $1$ second, since it deliberately postpones execution to the latest feasible instant. Furthermore, EDF-EG exhibits slightly higher latency in the low EH regime because the adaptive threshold mechanism introduced by OSTB increases the opportunities for harvesting additional energy, thereby enabling tasks to be executed earlier in the subsequent main interval.

\subsection{Performance under correlated EH}
\begin{figure}
  \centering
  \includegraphics[width=0.48\textwidth]{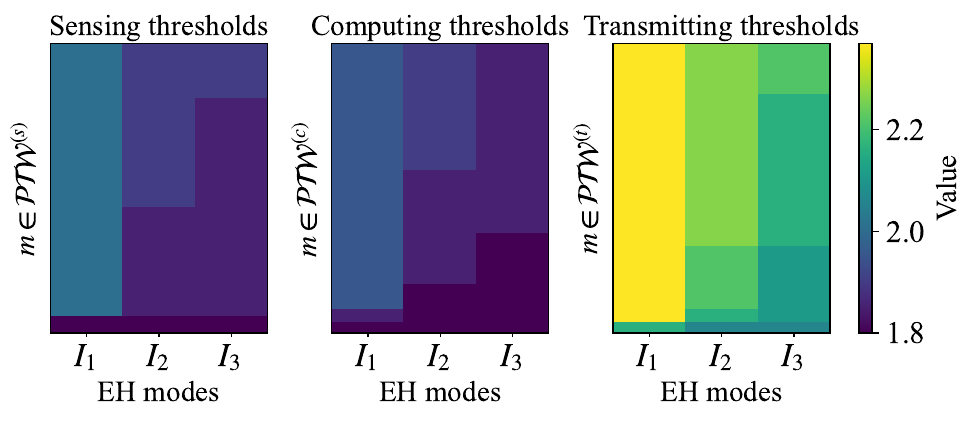}
  \caption{Threshold values as functions of within-frame position (i.e., $m$) for different tasks under different EH modes. The harvested-current levels are $\{I_h\}_{h=1}^{3} = [0.01,\ 1.5,\ 2.8]$mA, and the capacitance is $C = 1.7$mF.}
  \label{fig:Threshold_for_correlated_case}
\end{figure}
To evaluate the correlated-EH extension, we consider a finite-state Markov EH model with $L=3$ EH modes. We assume that the corresponding harvested-current levels are denoted by $\{I_h\}_{h=1}^3$ mA, and the mode evolution is governed by the transition matrix
\[
\mathbf P_H=\begin{bmatrix}
0.86 & 0.12 & 0.02 \\ 0.10 & 0.80 & 0.10 \\ 0.03 & 0.20 & 0.77
\end{bmatrix}.
\] 
Across all tested correlated-EH configurations, the optimal augmented policy remains threshold-based in the voltage variable within each admissible superstate, but the thresholds now depend explicitly on the EH mode. As shown in Fig.~\ref{fig:Threshold_for_correlated_case}, for fixed within-frame position, the threshold associated with a higher EH mode is lower than that associated with a lower mode. This ordering is especially visible for the transmitting task, which is the most energy-demanding stage of the chain. Hence, correlation in the EH process does not destroy the OSTB structure and instead enhances it by introducing an additional modal dimension. 
In addition, the numerical monotonicity check of the action-advantage functions confirms the structural claim above. Specifically, the computed quantities $\Delta_s(V,m,h)$, $\Delta_c(V,m,h)$, and $\Delta_t(V,m,h)$ are non-decreasing in $V$ up to the adopted numerical tolerance, which directly validates the threshold interpretation of the optimal policy in the discretized correlated-EH model.

\begin{figure}
  \centering
  \includegraphics[width=0.29\textwidth]{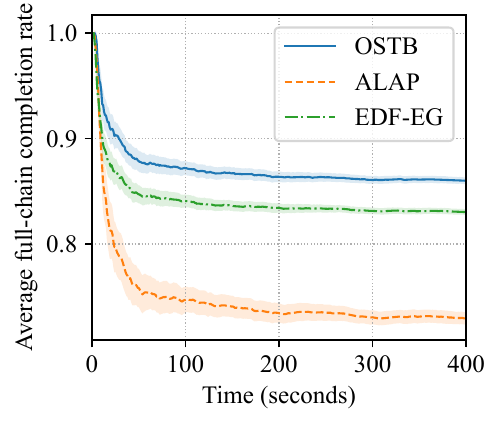}
  \caption{Average full-chain completion rate as functions of time for the proposed policy and the two baseline schemes. The correlated EH process is characterized by three harvested-current modes, $\{I_h\}_{h=1}^{3} = [0.01,\ 1.5,\ 2.8]$mA, and the capacitance is $C = 1.7$mF. The shaded areas indicate the spread obtained from $100$ Monte Carlo simulation runs.}
  \label{fig:correlated_case_different_policies}
\end{figure}
Fig.~\ref{fig:correlated_case_different_policies} highlights the performance advantage of the proposed OSTB policy in the correlated EH setting. The figure shows the average full-chain completion rate, that is, the fraction of cycles in which all three tasks are successfully executed. OSTB consistently outperforms the two baseline schemes. This gain stems from its adaptive threshold-based decision structure, which exploits the correlation in the EH process and adjusts task initiation according to both the available capacitor energy and the expected short-term energy conditions. As a result, the proposed policy is able to schedule tasks more efficiently, improving end-to-end cycle completion.

\section{Conclusion and Future Directions}
\label{sec:conclusion}
This paper investigated long-term energy-aware task scheduling for battery-less IoT devices operating under intermittent EH and hard timing constraints. We considered a periodic sensing--computing--transmitting task chain in which each task must be completed safely despite stochastic energy arrivals. To address this problem, we developed an MDP-based framework that explicitly captures capacitor-voltage dynamics, task precedence, and permissible execution windows, and formulated the objective as the maximization of the long-term average reward. We proved that the resulting MDP is unichain and further established that the optimal stationary policy has a threshold-based structure, which leads to the proposed OSTB scheduler. In addition, we introduced a sigmoid-based reward design to better shape the tradeoff between task progress and identified sufficient conditions under which the OSTB scheduler achieves optimality in correlated EH scenarios.
The numerical results demonstrate that OSTB achieves the largest number of successful full-chain epochs across the considered EH distributions, and reduces latency relative to baselines in low-energy conditions while maintaining end-to-end completion performance. All in all, these findings confirm that explicitly coupling energy availability with deadline urgency through an adaptive threshold policy is highly beneficial for battery-less periodic operation.

Several limitations remain for future work. First, OSTB is model-based, and its performance depends on the accuracy of the MDP model, including capacitor dynamics, load current, discretization, and EH statistics. Second, finer voltage and EH discretization improve accuracy but increase offline computation and memory cost, motivating approximate or hierarchical methods for larger state spaces. Third, extending the threshold analysis beyond periodic task chains to heterogeneous task graphs with richer dependencies requires further study. Fourth, non-stationary EH conditions may require online model estimation or adaptive EH-state transition models. Finally, hardware or hardware-in-the-loop validation is needed to quantify sensing overhead, voltage-measurement error, PMU inefficiencies, and implementation latency.

\appendices
\section{Proof of Theorem \ref{th:unichain}}
\label{app:th1}
We first introduce a higher-level representation of the chain.  
Given deterministic stationary policy~$\pi$, define the \emph{policy-induced superstate-level graph} $G_\pi = (R, E_\pi)$, where
\begin{itemize}
    \item the vertex set $R$ consists of all superstates $(\tau, f)$; and
    \item there is a directed edge $S \to T$ if, under $\pi$, some state in $S = (\tau_1, f_1)$ has positive probability of transitioning to $T = (\tau_2, f_2)$ in one step at the superstate level.
\end{itemize}
Because matrices $A_{l}$, $A_s$, $A_c$, and $A_t$ have strictly positive entries, the following holds.
If $S \to T$ in~$G_\pi$, then there exists at least one state in $S$ that has positive probability of reaching \emph{every} $N_v$-component of $T$ in one superstate-level step.  
Thus, studying reachability and recurrence at the superstate-level is valid for the full state-level chain.

\textbf{Tail superstates:}  
We define the \emph{tail superstates} as
\begin{align}
(\tau, f=0), & \quad \tau \in \{d_s+1, \dots, M-1\}, \nonumber\\
(\tau, f=1), & \quad \tau \in \{M-n_{c}-n_{t}+1, \dots, M-1\}, \nonumber\\
(\tau, f=2), & \quad \tau \in \{M-n_{t}+1, \dots, M-1\}, \nonumber\\
(\tau, f=3), & \quad \tau \in \{n_s+n_{c}+n_{t}, \dots, M-1\}.\nonumber
\end{align}
In these superstates, the only allowable action is \texttt{sleeping}, which advances~$\tau$ by $+1$ without changing~$f$ until reaching $M-1$, at which point the process wraps to $(0, 0)$.

\textbf{Reachability of $(0,0)$:}  
From any initial state $(V^{(i)}, \tau, f)$, the chain reaches some tail superstate under any $\pi$.  
Once in a tail superstate, repeatedly chosen \texttt{sleeping} action advances~$\tau$ until~$\tau=M-1$ and then moves to $(0,0)$.  
Therefore, $(0,0)$ is reachable from every state in the chain with strictly positive probability.

\textbf{Possible disconnections in $G_\pi$ and transience:}  
Two cases may lead to transient states:
\begin{enumerate}
    \item \emph{Disconnection at} $(\tau_0,f=0) \to (\tau_0+1,f=0)$, for $\tau_0 \in \mathcal{PTW}^{(s)}$.  
    This can occur if the agent chooses \texttt{sensing} for \emph{all} states in $(\tau_0,f=0)$.  
    In this case, all subsequent tail superstates $(\tau, f=0)$ with $\tau>\tau_0$ become transient.
    \item \emph{Disconnection at} $(\tau_1,f=1) \to (\tau_1+1,f=1)$, for $\tau_1 \in \mathcal{PTW}^{(c)}$.
    This can occur if the agent chooses \texttt{computing} for \emph{all} states in $(\tau_1,f=1)$.
    In this case, all subsequent tail superstates $(\tau, f=1)$ with $\tau>\tau_1$ are transient.
    \item \emph{Disconnection at} $(\tau_2,f=2) \to (\tau_2+1,f=2)$, for $\tau_2 \in \mathcal{PTW}^{(t)}$.
    This can occur if the agent chooses \texttt{transmitting} for \emph{all} states in $(\tau_2,f=2)$.
    In this case, all subsequent tail superstates $(\tau, f=2)$ with $\tau>\tau_2$ are transient.
\end{enumerate}

\textbf{Existence of exactly one recurrent class:}  
Despite these possible transient sets, note that
\begin{itemize}
    \item every state in the chain has a positive-probability path to $(0,0)$ (reachability);
    \item once in $(0,0)$, the positivity of the micro transition matrices ensures communication between all $N_v$-components in every superstate reachable from $(0,0)$;
    \item any recurrent class must contain $(0,0)$, because $(0,0)$ is reachable from all states and recurrent classes are closed.
\end{itemize}
Since there can be at most one recurrent class containing $(0,0)$, the MDP is unichain.
\hfill $\square$

\section{Proof of Theorem \ref{th:threshold}}
\label{app:th2}
\begin{proof}
Fix an arbitrary index $m \in \mathcal{PTW}^{(s)}$, and consider the ordered set of states in superstate $(m,0)$. Since the considered MDP is finite and admits an optimal stationary policy under the long-term average reward criterion, it is sufficient to verify the assumptions of \cite[Th.~8.11.3]{puterman1994} on each such superstate. We therefore show that, for fixed $m$, both the reward function and the transition matrix satisfy the required monotonicity and superadditivity properties.

\medskip
\noindent
1) Monotonicity and superadditivity of the reward.
From the definition of the reward in \eqref{eq:rewarddefinn2} or \eqref{eq:rewarddefin2}, we have $r((V,m,0),\texttt{sleeping})=0$, whereas $r((V,m,0),\texttt{sensing})$
is a non-decreasing function of $V$. Indeed, the safe-execution probability of the sensing task is non-decreasing in the initial capacitor voltage, and the reward is defined as a non-decreasing transformation of that probability. Hence, for any $V' \geq V$,
\[
r((V',m,0),\texttt{sensing}) \geq r((V,m,0),\texttt{sensing}),
\]
while the reward under \texttt{sleeping} remains zero. Therefore,
\begin{align}
&r((V',m,0),\texttt{sensing})-r((V',m,0),\texttt{sleeping})\nonumber\\
&\geq r((V,m,0),\texttt{sensing}) - r((V,m,0),\texttt{sleeping}),\nonumber
\end{align}
which is precisely the increasing-differences property in $(V,a)$. Since the action set is binary and ordered, this is equivalent to superadditivity of the reward, i.e., for $V' \geq V$
\begin{align}
&r((V',m,0),\texttt{sensing}) + r((V,m,0),\texttt{sleeping}) \nonumber\\
&\geq r((V',m,0),\texttt{sleeping}) + r((V,m,0),\texttt{sensing}).\nonumber
\label{eq:reward_superadditivity_appendix}
\end{align}

\medskip
\noindent
2) Monotonicity and superadditivity of the transition matrix.
Let $A_l$ and $A_s$ denote the transition matrices associated with the actions \texttt{sleeping} and \texttt{sensing}, respectively. By construction of the voltage dynamics, a larger initial capacitor voltage yields a stochastically larger distribution for the next voltage state, under either action. Hence both $A_l$ and $A_s$ are stochastically monotone with respect to $V$.
Equivalently, for every bounded non-decreasing function $\varphi$ on the next-state space and for any $V' \geq V$,
\[ \sum_j A_a(i',j)\,\varphi(V^{(j)}) \geq \sum_j A_a(i,j)\,\varphi(V^{(j)}), \]
where $V^{(i')}=V'$ and $V^{(i)}=V$.
Moreover, the transition structure satisfies the superadditivity requirement of \cite[Th.~8.11.3]{puterman1994}, namely that the map
\[ (V,a)\mapsto \sum_j A_a(i,j)\,\varphi(V^{(j)}) \]
has increasing differences in $(V,a)$ for every bounded non-decreasing $\varphi$. Therefore, the transition matrix is stochastically superadditive in the ordered pair $(V,a)$.

Having verified the structural assumptions of \cite[Th. 8.11.3]{puterman1994}, we conclude that the optimal stationary policy $\pi^*$ exhibits a threshold-based structure in $V$. The proof for superstates $(m,f=1)_{m\in\mathcal{PTW}^{(c)}}$ and $(m,f=2)_{m\in\mathcal{PTW}^{(t)}}$ is the same.
\end{proof}

\section{Proof of Theorem \ref{theorem:non_iid}}
\label{app:non_iid}
The proof follows the same monotone dynamic programming logic used in the i.i.d. setting, with the key distinction that the EH mode is now treated as piece-wise fixed within each augmented superstate. In particular, each augmented superstate $(\tau,f)$ is partitioned into $L$ fixed-mode superstates of the form $(\tau,f,h), \forall h\in \mathcal{H}$.
Since each admissible superstate admits only two naturally ordered actions, the monotonicity of the optimal policy can be established by analyzing the ordering of these two actions.
Therefore, it is sufficient to show that the difference between the corresponding state-action value functions is non-decreasing in $V$.


Let $b(s)$ denote a differential value function associated with the optimal average reward $g$, and define the stage-dependent action-advantage function
\begin{align}
\Delta_i(s) \triangleq
&\Big(r(s,i)-g\,d(i)+\sum_{s'}P(s'\mid s,i)b(s')\Big)\nonumber\\
&-\Big(r(s,l)-g\,d(l)+\sum_{s'}P(s'\mid s,l)b(s')\Big),
\end{align}
where $i\in\mathcal{I}$ is the unique executable task in the considered fixed-mode superstate. Note that the immediate reward difference between the executable action and \texttt{sleeping} is non-decreasing in $V$, and under all actions, the transition matrix is stochastically monotone in $V$.
Under these considerations and assumption (i), the dynamic-programming operator preserves monotonicity and increasing differences in the voltage variable. Consequently, $\Delta_i(s)$ is non-decreasing in $V$ for every fixed $(\tau,f,h), \forall h\in\mathcal{H}$. Therefore, the set
\[
\{V:\Delta_i(V,\tau,f,h)\ge 0\}
\]
is an upper set in $V$. This proves the existence of a threshold-based optimization policy. Furthermore, since the MDP is unichain by assumption (ii), the optimal policy is stationary under the average-reward criterion.
\hfill $\square$

\bibliographystyle{IEEEtran}
\bibliography{reference.bib}

@article{moser2007realtime,
author = {Moser, Clemens and others},
title = {{Real-time scheduling for energy harvesting sensor nodes}},
year = {2007},
issue_date = {December  2007},
publisher = {Kluwer Academic Publishers},
address = {USA},
volume = {37},
number = {3},
issn = {0922-6443},
doi = {10.1007/s11241-007-9027-0},
journal = {Real-Time Syst.},
month = dec,
pages = {233–260},
numpages = {28},
}

@misc{onelnewpaper,
      title={Foundations for Energy-Aware Zero-Energy Devices: From Energy Sensing to Adaptive Protocols}, 
      author={Onel L. A. López and others},
      year={2025},
      eprint={2507.22740},
      archivePrefix={arXiv},
      primaryClass={eess.SY},
      journal={arxiv}
}

@ARTICLE{delgado2022optimal,
  author={Delgado, Carmen and Famaey, Jeroen},
  journal = {IEEE Trans. Emerg. Top. Comput.},
  title={Optimal Energy-Aware Task Scheduling for Batteryless IoT Devices}, 
  year={2022},
  volume={10},
  number={3},
  pages={1374-1387},
  doi={10.1109/TETC.2021.3086144}}

@book{puterman1994,
  author    = {Martin Lee Puterman},
  title     = {{Markov Decision Processes: Discrete Stochastic Dynamic Programming}},
  publisher = {Wiley},
  year      = {2005},
  address   = {US},
  edition   = {Second},
  isbn      = {9780471619772},}

@article{biason2018decentralized,
author = {Biason, Alessandro and others},
title = {A Decentralized Optimization Framework for Energy Harvesting Devices},
year = {2018},
issue_date = {Nov. 2018},
publisher = {IEEE Educational Activities Department},
address = {USA},
volume = {17},
number = {11},
issn = {1536-1233},
doi = {10.1109/TMC.2018.2810269},
journal = {IEEE Trans. Mobile Comput.},
month = nov,
pages = {2483–2496},
numpages = {14}
}

@Article{sabovic2020lora,
 AUTHOR = {Sabovic, Adnan and others},
 TITLE = {Energy-Aware Sensing on Battery-Less {LoRaWAN} Devices with Energy Harvesting},
 JOURNAL = {Electronics},
 VOLUME = {9},
 YEAR = {2020},
 NUMBER = {6},
 ARTICLE-NUMBER = {904},
 ISSN = {2079-9292},
 DOI = {10.3390/electronics9060904}}

@ARTICLE{sabovic2022scheduler,
  author={Sabovic, Adnan and others},
  journal = {IEEE Internet Things J.}, 
  title={An Energy-Aware Task Scheduler for Energy-Harvesting Batteryless {IoT} Devices}, 
  year={2022},
  volume={9},
  number={22},
  pages={23097-23114},
  doi={10.1109/JIOT.2022.3185321}
  }

@misc{rioual2021design,
      title={Design and Comparison of Reward Functions in Reinforcement Learning for Energy Management of Sensor Nodes}, 
      author={Yohann Rioual and Yannick Le Moullec and Johann Laurent and Muhidul Islam Khan and Jean-Philippe Diguet},
      year={2021},
      eprint={2106.01114},
      archivePrefix={arXiv},
      primaryClass={eess.SY},
}

@inproceedings{Ghor2025DeadlineDriven,
  author    = {Mohamad El Ghor and Maryline Chetto and Hussein El Ghor},
  title     = {Deadline-Driven Harvesting-aware Scheduling for Autonomous Cyber-Physical Systems},
  booktitle = {8th IEEE CiSt},
  pages     = {310--315},
  year      = {2025},
  publisher = {IEEE},
  doi       = {10.1109/CiSt65886.2025.11224132}
}

@ARTICLE{9174941,
  author={Delgado, Carmen and others},
  journal = {IEEE Internet Things J.}, 
  title={Batteryless {LoRaWAN} Communications Using Energy Harvesting: Modeling and Characterization}, 
  year={2021},
  volume={8},
  number={4},
  pages={2694-2711},
  doi={10.1109/JIOT.2020.3019140}}

@ARTICLE{sustainable,
  author={López, Onel L. A. and others},
  journal={IEEE Open J. Commun. Soc.}, 
  title={Energy-Sustainable {IoT} Connectivity: Vision, Technological Enablers, Challenges, and Future Directions}, 
  year={2023},
  volume={4},
  number={},
  pages={2609-2666},
  doi={10.1109/OJCOMS.2023.3323832}
  }

@ARTICLE{kang2024powerdepletion,
AUTHOR = {Kang, Young-myoung and Lim, Yeon-sup},
TITLE = {Handling Power Depletion in Energy Harvesting {IoT} Devices},
JOURNAL = {Electronics},
VOLUME = {13},
YEAR = {2024},
NUMBER = {14},
ARTICLE-NUMBER = {2704},
ISSN = {2079-9292},
DOI = {10.3390/electronics13142704}
}

@misc{jahanbazi2026multiexit,
      title={Energy-Aware Multi-Exit {TinyML} for Smart Zero-Energy Devices}, 
      author={Shahab Jahanbazi and others},
      year={2026},
      eprint={2603.08047},
      archivePrefix={arXiv},
      primaryClass={eess.SP},
}

@INPROCEEDINGS{8715130,
  author={Singla, Priyanka and others},
  booktitle={2019 Design, Automation \& Test in Europe Conference \& Exhibition (DATE)}, 
  title={{FlexiCheck}: An Adaptive Checkpointing Architecture for Energy Harvesting Devices}, 
  year={2019},
  volume={},
  number={},
  pages={546-551},
  doi={10.23919/DATE.2019.8715130}}

@ARTICLE{8107579,
  author={Sakulkar, Pranav and Krishnamachari, Bhaskar},
  journal = {IEEE Trans. Inf. Theory},
  title={Online Learning Schemes for Power Allocation in Energy Harvesting Communications}, 
  year={2018},
  volume={64},
  number={6},
  pages={4610-4628},
  doi={10.1109/TIT.2017.2773526}}

@INPROCEEDINGS{rao2015optimal,
  author={Rao, Vijay S. and others},
  booktitle = {Proc. IEEE Wireless Commun. Netw. Conf. (WCNC)},
  title={{Optimal task scheduling policy in energy harvesting wireless sensor networks}}, 
  year={2015},
  volume={},
  number={},
  pages={1030-1035},
  doi={10.1109/WCNC.2015.7127611}}

@Article{periodic_sensing,
AUTHOR = {Peng, Cheng-Sheng and Wang, Chao},
TITLE = {Adaptive Transmissions for Batteryless Periodic Sensing},
JOURNAL = {IoT},
VOLUME = {5},
YEAR = {2024},
NUMBER = {2},
PAGES = {332--355},
ISSN = {2624-831X},
DOI = {10.3390/iot5020017}
}

@Article{hao2019robust,
AUTHOR = {Hao, Jie and others},
TITLE = {A Robust Transmission Scheduling Approach for Internet of Things Sensing Service with Energy Harvesting},
JOURNAL = {Sensors},
VOLUME = {19},
YEAR = {2019},
NUMBER = {14},
ARTICLE-NUMBER = {3090},
PubMedID = {31336962},
ISSN = {1424-8220},
DOI = {10.3390/s19143090}
}

@ARTICLE{9474499,
  author={Prasad, R. Venkatesha and others},
  journal = {IEEE Trans. Green Commun. Netw.}, 
  title={{ReNEW: A Practical Module for Reliable Routing in Networks of Energy-Harvesting Wireless Sensors}}, 
  year={2021},
  volume={5},
  number={3},
  pages={1558-1569},
  doi={10.1109/TGCN.2021.3094771}}

@INPROCEEDINGS{Hicks2017clank,
  author={Hicks, Matthew},
  booktitle={2017 ACM/IEEE 44th Annual International Symposium on Computer Architecture (ISCA)}, 
  title={{Clank: Architectural support for intermittent computation}}, 
  year={2017},
  volume={},
  number={},
  pages={228-240},
  doi={10.1145/3079856.3080238}}

@book{ross,
  author    = {Sheldon M. Ross},
  title     = {Introduction to Stochastic Dynamic Programming},
  publisher = {Academic},
  year      = {1983},
  address   = {CA, USA},
  doi     = {10.1016/C2013-0-11415-8},}

@INPROCEEDINGS{islam2020deadlineaware,
  author={Islam, Bashima and Nirjon, Shahriar},
  booktitle={2020 IEEE Real-Time and Embedded Technology and Applications Symposium (RTAS)}, 
  title={Scheduling Computational and Energy Harvesting Tasks in Deadline-Aware Intermittent Systems}, 
  year={2020},
  volume={},
  number={},
  pages={95-109},
  doi={10.1109/RTAS48715.2020.00-14}}

@ARTICLE{karimi2021realtime,
  author={Karimi, Mohsen and others},
  journal = {IEEE Internet Things J.}, 
  title={Real-Time Task Scheduling on Intermittently Powered Batteryless Devices}, 
  year={2021},
  volume={8},
  number={17},
  pages={13328-13342},
  doi={10.1109/JIOT.2021.3065947}}

@INPROCEEDINGS{karimi2022energyadaptive,
  author={Karimi, Mohsen and others},
  booktitle = {Proc. IEEE Int. Conf. Embedded Real-Time Comput. Syst. Appl. (RTCSA)},
  title={Energy-Adaptive Real-time Sensing for Batteryless Devices}, 
  year={2022},
  volume={},
  number={},
  pages={205-211},
  doi={10.1109/RTCSA55878.2022.00028}}

@ARTICLE{karimi2026cartos,
author={Karimi, Mohsen and others},
journal = {IEEE Trans. Emerg. Top. Comput.},
title={{ CARTOS: A Charging-Aware Real-Time Operating System for Intermittent Batteryless Devices }},
year={2026},
volume={14},
number={01},
ISSN={2168-6750},
pages={54-68},
doi={10.1109/TETC.2025.3643888},
publisher={IEEE Computer Society},
address={Los Alamitos, CA, USA},
month=jan}

@article{sabovic2023tinyml,
title = {Towards energy-aware {tinyML} on battery-less {IoT} devices},
journal = {Internet of Things},
volume = {22},
pages = {100736},
year = {2023},
issn = {2542-6605},
doi = {https://doi.org/10.1016/j.iot.2023.100736},
author = {Adnan Sabovic and others},
}

@inproceedings{fraternali2020ember,
author = {Fraternali, Francesco and others},
title = {{Ember}: energy management of batteryless event detection sensors with deep reinforcement learning},
year = {2020},
isbn = {9781450375900},
publisher = {Association for Computing Machinery},
address = {New York, NY, USA},
doi = {10.1145/3384419.3430734},
booktitle = {Proc. 18th Conf. Embedded Netw. Sensor Syst.},
pages = {503–516},
numpages = {14},
location = {Virtual Event, Japan},
series = {SenSys '20}
}

@article{fraternali2020aces,
author = {Fraternali, Francesco and others},
title = {{ACES}: Automatic Configuration of Energy Harvesting Sensors with Reinforcement Learning},
year = {2020},
issue_date = {November 2020},
publisher = {Association for Computing Machinery},
address = {New York, NY, USA},
volume = {16},
number = {4},
issn = {1550-4859},
doi = {10.1145/3404191},
month = jul,
articleno = {36},
numpages = {31},
}

@INPROCEEDINGS{luo2021smarton,
  author={Luo, Yubo and Nirjon, Shahriar},
  booktitle={2021 17th International Conference on Distributed Computing in Sensor Systems (DCOSS)}, 
  title={{SmartON}: {Just-in-Time} Active Event Detection on Energy Harvesting Systems}, 
  year={2021},
  volume={},
  number={},
  pages={35-44},

  doi={10.1109/DCOSS52077.2021.00018}}

@Article{Chetto2023EDF,
AUTHOR = {Chetto, Maryline and El Osta, Rola},
TITLE = {Earliest Deadline First Scheduling for Real-Time Computing in Sustainable Sensors},
JOURNAL = {Sustainability},
VOLUME = {15},
YEAR = {2023},
NUMBER = {5},
ARTICLE-NUMBER = {3972},
ISSN = {2071-1050},
DOI = {10.3390/su15053972}
}

@inproceedings{age-based10.1145/3209582.3209602,
author = {Lu, Ning and Ji, Bo and Li, Bin},
title = {Age-based Scheduling: Improving Data Freshness for Wireless Real-Time Traffic},
year = {2018},
isbn = {9781450357708},
publisher = {Association for Computing Machinery},
address = {New York, NY, USA},
doi = {10.1145/3209582.3209602},
pages = {191–200},
numpages = {10},
location = {Los Angeles, CA, USA},
series = {Mobihoc '18}
}

@article{KIM2025107542,
title = {A priority-aware dynamic scheduling algorithm for ensuring data freshness in {5G} networks},
journal = {Future Generation Computer Systems},
volume = {163},
pages = {107542},
year = {2025},
issn = {0167-739X},
author = {Beom-Su Kim},
}

@ARTICLE{5934952mobile1,
  author={Gorlatova, Maria and Wallwater, Aya and Zussman, Gil},
  journal = {IEEE Trans. Mobile Comput.}, 
  title={Networking Low-Power Energy Harvesting Devices: Measurements and Algorithms}, 
  year={2013},
  volume={12},
  number={9},
  pages={1853-1865},
  doi={10.1109/TMC.2012.154}}

\end{document}